\documentclass[usenatbib]{mn2e}
\usepackage{graphicx}
\usepackage{astrobib_mnras2e}

\begin{document}
\topmargin-1cm

\newcommand\approxgt{\mbox{$^{>}\hspace{-0.24cm}_{\sim}$}}
\newcommand\approxlt{\mbox{$^{<}\hspace{-0.24cm}_{\sim}$}}
\newcommand{\be}{\begin{equation}}
\newcommand{\ee}{\end{equation}}
\newcommand{\bea}{\begin{eqnarray}}
\newcommand{\eea}{\end{eqnarray}}
\newcommand{\lexp}{\mathop{\langle}}
\newcommand{\rexp}{\mathop{\rangle}}
\newcommand{\rexpc}{\mathop{\rangle_c}}
\newcommand{\pcl}{pseudo-$C_{l}$~}
\newcommand{\pclc}{Pseudo-$C_{l}$~}
\newcommand{\ylm}{Y_{lm}}
\newcommand{\plm}{\psi_{lm}}
\newcommand{\thi}{\hat{\theta}_{i}}
\newcommand{\thj}{\hat{\theta}_{j}}
\newcommand{\nbar}{\bar{n}}
\newcommand{\talm}{\tilde{a}_{lm}}
\newcommand{\tcl}{\tilde{C}_{l}}
\newcommand{\nhat}{\hat{\bf n}}
\newcommand{\kvec}{{\bf k}}
\newcommand{\khat}{\hat{\bf k}}
\def\bi#1{\hbox{\boldmath{$#1$}}}

\title{The Clustering of Luminous Red Galaxies in the
Sloan Digital Sky Survey Imaging Data}

\author[Padmanabhan et al]
{Nikhil Padmanabhan$^{1}$\thanks{npadmana@princeton.edu}, 
David J. Schlegel$^{2}$, 
Uro\v{s} Seljak$^{1,3}$, 
Alexey Makarov$^{1}$,\and
Neta A. Bahcall$^{4}$, 
Michael R. Blanton$^{5}$,
Jonathan Brinkmann$^{6}$,
Daniel J. Eisenstein$^{7}$,\and
Douglas P. Finkbeiner$^{4}$,
James E. Gunn$^{4}$, 
David W. Hogg$^{5}$,
\v{Z}eljko Ivezi\'{c}$^{8}$,\and
Gillian R. Knapp$^{4}$,
Jon Loveday$^{9}$,
Robert H. Lupton$^{4}$, 
Robert C. Nichol$^{10}$, \and
Donald P. Schneider$^{11}$, 
Michael A. Strauss$^{4}$,
Max Tegmark$^{12}$,
Donald G. York$^{13}$
\\
$^{1}$Joseph Henry Laboratories, Jadwin Hall, Princeton University, Princeton, NJ 08544, USA \\
$^{2}$Physics Division, Lawrence Berkeley National Laboratories, Berkeley, CA 94720-8160, USA \\
$^{3}$ICTP, Strada Costiera 11, 34014 Trieste, Italy \\
$^{4}$Dept. of Astrophysical Sciences, Peyton Hall, Princeton University, Princeton, NJ 08544, USA \\ 
$^{5}$Department of Physics, New York University, 4 Washington Pl, New York, NY 10003, USA \\
$^{6}$Apache Point Observatory, 2001 Apache Point Road, Sunspot, NM 88349-0059, USA \\
$^{7}$Steward Observatory, University of Arizona, 933 N. Cherry Ave, Tucson, AZ 85721, USA \\
$^{8}$University of Washington, Department of Astronomy, Box 351580, Seattle, WA 98195, USA \\
$^{9}$Astronomy Centre, University of Sussex, Falmer, Brighton, BN1 9QH, UK \\
$^{10}$Institute of Cosmology and Gravitation, University of Portsmouth, Portsmouth, Portsmouth, PO12EG, UK \\
$^{11}$Department of Astronomy \& Astrophysics, The Pennsylvania 
State University,525 Davey Laboratory, University Park, PA 16802, USA \\
$^{12}$MIT Kavli Institute for Astrophysics and Space Research, Cambridge, MA 02139, USA \\
$^{13}$Dept. of Astronomy and Astrophysics,\& Enrico Fermi Institute, 5640 So. Ellis Avenue, Chicago, IL 60637, USA\\
}

\date{\today}
\maketitle

\begin{abstract}
We present the 3D real space clustering power spectrum of a sample of 
$\sim 600,000$ luminous red galaxies (LRGs) measured by the Sloan Digital Sky 
Survey (SDSS), using photometric redshifts. These galaxies are old, elliptical 
systems with strong $4000$~\AA\, breaks, and have accurate photometric redshifts
with an average error of $\Delta z = 0.03$. This sample of galaxies ranges from  redshift
$z=0.2$ to $0.6$ over $3,528\, {\rm deg}^{2}$ of the sky, probing a volume of 
$1.5 h^{-3} {\rm Gpc}^{3}$, making it the largest volume ever used for galaxy clustering
measurements. We measure the angular clustering power spectrum in eight redshift slices
and use well-calibrated redshift distributions to combine these into 
a high precision 3D real space power spectrum from $k=0.005 h {\rm Mpc}^{-1}$ to $k=1 h {\rm Mpc}^{-1}$. 
We detect power on gigaparsec scales, beyond the turnover in the matter power spectrum,
at a $\sim 2 \sigma$ significance for $k < 0.01 h {\rm Mpc}^{-1}$, increasing to $5.5 \sigma$
for $k < 0.02 h {\rm Mpc}^{-1}$. This detection of power is on scales significantly larger
than those accessible to current spectroscopic redshift surveys. 
We also find evidence for baryonic oscillations, both in the power spectrum, as well as in 
fits to the baryon density, at a $2.5 \sigma$ confidence level. The large volume and resulting
small statistical errors on the power spectrum allow us to constrain both the amplitude and
scale dependence of the galaxy bias in cosmological fits. The statistical power of these data
to constrain cosmology is $\sim 1.7$ times better than previous clustering analyses.
Varying the matter density and 
baryon fraction, we find $\Omega_{M} = 0.30 \pm 0.03$, and $\Omega_{b}/\Omega_{M} = 0.18 \pm 0.04$,
for a fixed Hubble constant of $70\,{\rm km/s/Mpc}$ and a scale-invariant spectrum 
of initial perturbations. The detection of baryonic oscillations also allows us to
measure the comoving distance to $z=0.5$; we find a best fit distance of $1.73 \pm 0.12 {\rm Gpc}$,
corresponding to a $6.5\%$ error on the distance. 
These results demonstrate the ability to make precise clustering measurements with photometric surveys.
\end{abstract}



\section{Introduction}

The three dimensional distribution of galaxies has long been recognized as
a powerful cosmological probe 
\citep{1973ApJ...185..413P, 1973ApJ...185..757H, 1977ApJ...217..385G,
1997PhRvL..79.3806T, 1998ApJ...499..555T,
1998ApJ...495...29G, 1998PhRvL..80.5255H, 1999ApJ...510...20W, 
1999PhRvD..59b3512H, 1999ApJ...518....2E}. 
On large scales, we expect galaxy density to have a simple relationship
to the underlying matter density;
therefore, the clustering of galaxies is related
to the clustering of the underlying matter. The two point correlation 
function of matter (or its Fourier transform, the power spectrum) is a sensitive
probe of both the initial conditions of the Universe and its subsequent 
evolution. Indeed, if the matter density is well described by a Gaussian 
random field, then the power spectrum encodes all the information present in
the field. It is therefore not surprising that a large fraction of the 
effort in observational cosmology has been devoted to measuring the spatial 
distribution of galaxies, culminating in recent results from the Two-Degree Field 
Galaxy Redshift Survey \citep[2dFGRS,][]{2005MNRAS.362..505C} 
and the Sloan Digital Sky Survey \citep[SDSS,][]{2004ApJ...606..702T}.

The spatial distribution of galaxies is also a standard 
ruler for cosmography. The expansion rate of the Universe as a function of redshift
is a sensitive probe of its energy content, and in particular, 
can be used to constrain the properties of the ``dark energy'' responsible for the
recent acceleration in the expansion \cite[see eg.][]{2005ASPC..339..215H, 2005NewAR..49..360E}. 
One approach to measure the expansion rate
is to observe the apparent size of a standard ruler (and therefore,
the angular diameter distance) at different redshifts to constrain
the scale factor as a function of time.
The power spectrum of the galaxy distribution has two features useful as standard 
rulers. At $k \sim 0.01 h \rm{Mpc}^{-1}$, the power spectrum turns over from a $k^{1}$ slope
(for a scale invariant spectrum of initial fluctuations), to a $k^{-3}$ spectrum, 
caused by modes that entered the horizon 
during radiation domination and were therefore suppressed. The precise
position of this turnover is determined by the size of the horizon 
at matter-radiation equality, and corresponds to a 
physical scale determined by the matter $(\Omega_{M}h^{2})$ and 
radiation densities $(\Omega_{R}h^{2})$. The other distinguishing feature is oscillations in
the power spectrum caused by acoustic waves in the baryon-photon plasma before hydrogen 
recombination at $z\sim1000$ 
\citep{1970ApJ...162..815P,1980ARA&A..18..537S, 1984ApJ...285L..45B, 
1989ApJS...71....1H, 1998ApJ...496..605E,1999MNRAS.304..851M}. 
The physics of these oscillations are analogous to those of
the cosmic microwave background, although their amplitude is suppressed 
because only $\sim 1/6$ of the matter in the Universe is composed of baryons.
The scale of this feature, again determined by the matter and radiation
densities, is set by the sound horizon at hydrogen recombination. This feature was 
first observed in early 2005 both in the SDSS  
Luminous Red Galaxy sample \citep{2005ApJ...633..560E, 2006A&A...449..891H}
and the 2dFGRS data \citep{2005MNRAS.362..505C}. 
Measuring the apparent size of both of these features at different 
redshifts opens up the possibility of directly measuring the angular 
diameter distance as a function of redshift \citep{1998ApJ...504L..57E, 2003ApJ...598..720S,
2003PhRvD..68h3504L, 2003PhRvL..90b1302M, 2003ApJ...594..665B, 2003PhRvD..68f3004H, 2004ApJ...615..573M,
2005ApJ...633..575S, 2005APh....24..334W,
2005MNRAS.363.1329B, 2006MNRAS.365..255B,2006MNRAS.366..884D}.

Traditionally, measurements of galaxy clustering rely on spectroscopic redshifts
to estimate distances to galaxies. Even with modern CCDs and high throughput
multi-fiber spectrographs, acquiring them is an expensive, time-consuming process compared
with just imaging the sky. For instance,
the SDSS spends about one-fifth of the time imaging the sky, and the rest
on spectroscopy. Furthermore, the ultimate accuracy of distance estimates 
from spectroscopy is limited by peculiar velocities of $\sim 1000 {\rm km/s}$, a 
significant mismatch with the intrinsic 
spectroscopic accuracy of $\sim 10 {\rm km/s}$. 

Large multi-band imaging surveys allow for the possibility of replacing
spectroscopic with photometric redshifts. The advantage is
relative efficiency of imaging over spectroscopy. Given a constant amount of 
telescope time, one can image both wider areas and deeper volumes than would be possible 
with spectroscopy, allowing one to probe both larger scales and larger volumes. 
Furthermore, the accuracy of photometric distance estimates \citep{2005MNRAS.359..237P},
$c\Delta z \sim 10,000 {\rm km/s}$ is more closely matched (although still not optimal)
to the intrinisic uncertainties in the distance-redshift relations.

One aim of this paper is to demonstrate the practicality of such an approach by
applying it to real data. 
We start with the five band imaging of the SDSS, and photometrically select a 
sample of luminous red galaxies; these galaxies have a strong 4000 \AA\, break in 
their spectral energy distributions, making uniform selection and accurate 
photometric redshifts possible. We then measure the angular clustering power spectrum 
as a function of redshift, and ``stack'' these individual 2D power spectra to 
obtain an estimate of the 3D clustering power spectrum. Using the photometric 
survey allows us to probe both larger scales and higher redshifts than is 
possible with the SDSS spectroscopic samples.

We pay special attention to the systematics unique to photometric surveys,
and develop techniques to test for these. Stellar contamination, variations in 
star-galaxy separation with seeing, uncertainties in Galactic extinction,
and variations in the photometric calibration
all can masquerade as large scale structure, making it essential to understand the 
extent of their contamination. Furthermore, stacking the angular power spectra
to measure the 3D clustering of galaxies requires testing our understanding of the
photometric redshifts and their errors.

The paper is organized as follows : Sec.~\ref{sec:sample} describes the construction
of the sample; Sec.~\ref{sec:angular} then discusses the measurement of the angular
power spectrum and the associated checks for systematics. These angular 
power spectra are then stacked to estimate the 3D power spectrum (Sec.~\ref{sec:3d}),
and preliminary cosmological parameters are estimated in Sec.~\ref{sec:cosmo}. We 
conclude in Sec.~\ref{sec:discuss}. Wherever not explicitly mentioned, we assume
a flat $\Lambda$CDM cosmology with $\Omega_{M}=0.3$, $\Omega_{b}=0.05$, $h=0.7$, 
a scale invariant primordial power spectrum, and $\sigma_{8}=0.9$.

\section{The Sample}
\label{sec:sample}

\subsection{The Data}

The Sloan Digital Sky Survey \citep{2000AJ....120.1579Y} is an
ongoing effort to image approximately $\pi$ steradians of the sky, and
obtain spectra of approximately one million of the detected objects
\citep{2002AJ....124.1810S, 2001AJ....122.2267E}. The
imaging is carried out by drift-scanning the sky in photometric conditions
\citep{2001AJ....122.2129H}, using a 2.5m telescope \citep{2006AJ....131.2332G}
in five bands ($ugriz$)
\citep{1996AJ....111.1748F, 2002AJ....123.2121S} using a specially designed
wide-field camera \citep{1998AJ....116.3040G}.
Using these data, objects are targeted for spectroscopy
\citep{2002AJ....123.2945R,2003AJ....125.2276B}
and are observed with a 640-fiber spectrograph on the same telescope. All
of these data are processed by completely automated pipelines that detect
and measure photometric properties of objects, and astrometrically
calibrate the data 
\citep{2001adass..10..269L, 2003AJ....125.1559P,2004AN....325..583I}. The
first phase of the SDSS is complete and has produced 
five major data releases
\citep{2002AJ....123..485S, 2003AJ....126.2081A, 2004AJ....128..502A,
2005AJ....129.1755A, 2006ApJS..162...38A}\footnote{URL: \texttt{www.sdss.org/dr4}}.
This paper uses all data observed through Fall 2003 (corresponding approximately
to SDSS Data Release 3), reduced as described by
\cite{2004AJ....128.2577F}.

\subsection{Photometric Calibration}

Measurements of large scale structure with a photometric survey require 
uniform photometric calibrations over the entire survey region. Traditional 
methods of calibrating imaging data involve comparisons with secondary ``standard''
stars. The precision of such methods is limited by transformations between 
different photometric systems, and there is no control over the relative photometry
over the entire survey region. The approach we adopt with these data is to use 
repeat observations of stars to constrain the photometric calibration of SDSS ``runs'',
analogous to CMB map-making techniques \citep[see eg.][]{1997ApJ...480L..87T}. 
Since all observations 
are made with the same telescope, there are none of the uncertainties associated with 
using auxiliary data. Also, using overlaps allows one to control the relative calibration
over connected regions of survey. The only uncertainty is the overall zeropoint of the 
survey, which we match to published SDSS calibrations. The above method has been briefly
described by \cite{2004AJ....128.2577F} and \cite{2005AJ....129.2562B}, 
and will be explained in detail in a future publication.

\subsection{Defining Luminous Red Galaxies}

Tracers of the large scale structure of the 
Universe must satisfy a number of criteria. They must probe a
large cosmological volume to overcome sample variance,
and have a high number density so shot noise
is sub-dominant on the scales of interest. Furthermore, it must 
be possible to uniformly select these galaxies over the entire 
volume of interest. Finally, if spectroscopic redshifts 
are unavailable, they should have well characterized photometric
redshifts (and errors), and redshift distributions.

\begin{figure}
\begin{center}
\leavevmode
\includegraphics[width=3.0in]{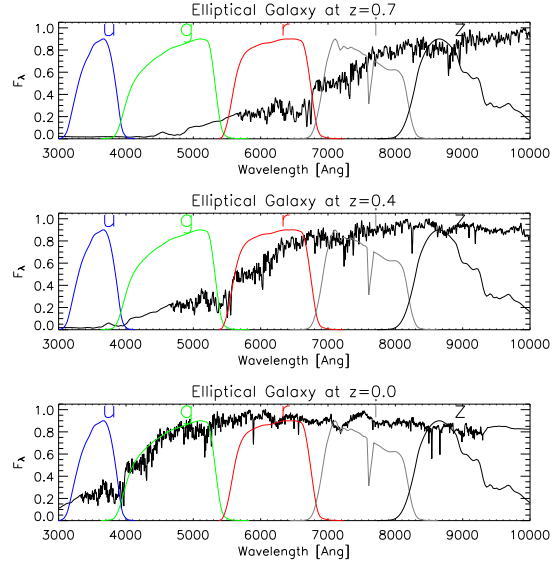}
\end{center}
\caption{A model spectrum of an elliptical galaxy, taken from 
Bruzual \& Charlot (\citeyear{2003MNRAS.344.1000B}), shown at 
three redshifts. The model assumes a single burst of star formation
11 Gyr ago and solar metallicity; the effect of evolution 
is not shown for simplicity. Also overplotted are the response
functions (including atmospheric absorption) for the five SDSS
filters. The break in the spectrum at 4000 \AA, and its migration
through the SDSS filters is clearly seen.
}
\label{fig:lrg_spectrum}
\end{figure}

The usefulness of LRGs as a cosmological
probe has long been appreciated
\citep{2000AJ....120.2148G, 2001AJ....122.2267E}. 
These are typically the most luminous galaxies in the Universe, 
and therefore probe cosmologically
interesting volumes.  In addition, these galaxies are generically old
stellar systems with uniform spectral energy distributions
(SEDs) characterized principally by a strong
discontinuity at 4000~\AA (Fig.~\ref{fig:lrg_spectrum}). 
This combination of a uniform SED and a
strong 4000~\AA~break make LRGs an ideal candidate for
photometric redshift algorithms, with redshift accuracies of $\sigma_z
\sim 0.03$ \citep{2005MNRAS.359..237P}. LRGs have been
used for a number of studies \citep{2004PhRvD..70j3501H,
2005ApJ...619..178E,2005ApJ...621...22Z, 2005PhRvD..72d3525P}, most 
notably for the detection of the baryonic acoustic peak in the galaxy 
autocorrelation function \citep{2005ApJ...633..560E}

\begin{figure}
\begin{center}
\leavevmode
\includegraphics[width=3.0in]{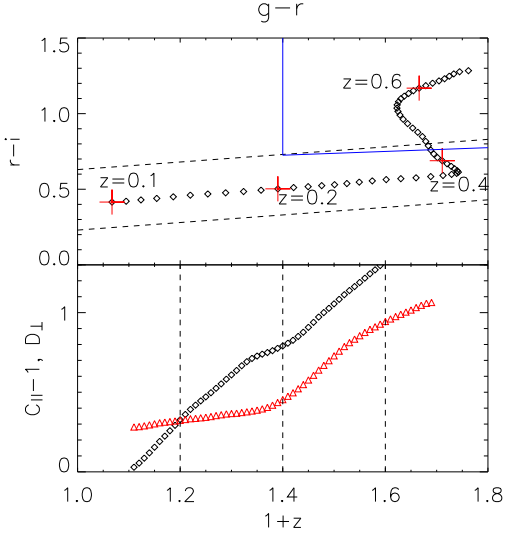}
\end{center}
\caption{The top panel shows simulated $g-r$ and $r-i$ colours of 
an early-type galaxy as
a function of redshift. The spectrum used to generate the 
track is the same as in Fig. \ref{fig:lrg_spectrum}, but evolved in 
redshift. Also shown are the colour cuts for Cut I (dashed, black)
and Cut II galaxies (solid, blue). The lower panel shows the 
colours $c_{||}$ (diamonds, black) and $d_{\perp}$ (triangles, red),
as a function of redshift. Also shown are fiducial redshift boundaries for
Cut I (0.2 -- 0.4) and Cut II (0.4 -- 0.6). Note that the range in $g-r$
is identical to the range in $1+z$.
}
\label{fig:lrg_colordiag}
\end{figure}

The photometric selection criteria we adopt were discussed in detail by 
\cite{2005MNRAS.359..237P} and are summarized below.
We start with a model spectrum of an early type galaxy 
from the stellar population synthesis models of \cite{2003MNRAS.344.1000B} (Fig.~\ref{fig:lrg_spectrum}).
This particular spectrum is derived from a single burst of star formation 11 Gyr ago
(implying a redshift of formation, $z_{form} \sim 2.6$), 
evolved to the present, and is typical of LRG spectra. In particular, the
4000 \AA~break is very prominent.
To motivate our selection criteria, we passively
evolve this spectrum in redshift
(taking the evolution of the strength of the 4000 \AA~break
into account), and project it through the SDSS filters; the 
resulting colour track in $g-r-i$ space as a function of redshift is shown in
Fig. \ref{fig:lrg_colordiag}. The bend in the track around $z \sim 0.4$, as 
the 4000 \AA~break redshifts from the $g$ to $r$ band, naturally 
suggests two selection criteria -- a low redshift sample (Cut I), nominally from
$z \sim 0.2 - 0.4$, and a high redshift sample (Cut II), from $z \sim 
0.4 - 0.6$. We define the two colours 
\citep[][and private commun.]{2001AJ....122.2267E}
\bea
c_{\perp} \equiv (r-i) - (g-r)/4 - 0.18 \,\,\, ,\\
d_{\perp} \equiv (r-i) - (g-r)/8 \approx r-i \,\,\,.
\label{eq:perpdef}
\eea
We now make the following colour selections,
\bea
{\rm Cut\,\,I :} & \mid c_{\perp} \mid < 0.2 \,\,\,;\\
{\rm Cut\,\,II :} & d_{\perp} > 0.55 \,\,\,, \\
& g-r > 1.4 \,\,\,,
\label{eq:colourcuts}
\eea
as shown in Fig. \ref{fig:lrg_colordiag}. The final cut, $g-r > 1.4$, 
isolates our sample from the stellar locus. In addition to these selection
criteria, we eliminate all galaxies 
with $g-r > 3$ and $r-i > 1.5$; these constraints
eliminate no real galaxies, but are effective at removing stars with 
unusual colours.

Unfortunately, as emphasized in \cite{2001AJ....122.2267E}, these simple colour
cuts are not sufficient to select LRGs due to an accidental degeneracy 
in the SDSS filters that causes all galaxies, irrespective 
of type, to lie very close to the low redshift early type locus. We therefore
follow the discussion there and impose a cut in absolute magnitude. We implement
this by defining a colour as a proxy for redshift and then translating
the absolute magnitude cut into a colour-apparent magnitude cut. We see from Fig. 
\ref{fig:lrg_colordiag} that $d_{\perp}$ correlates strongly with redshift and
is appropriate to use for Cut II. For Cut I, we define,
\be
c_{||} = 0.7 (g-r) + 1.2(r-i-0.18) \,\,\,,
\label{eq:cpllel}
\ee
which is approximately parallel to the low redshift locus. Given these, we 
further impose
\bea
{\rm Cut\,\,I :} & r_{Petro} < 13.6 + c_{||}/0.3 \,\,\,,\nonumber \\
& r_{Petro} < 19.7 \,\,\,; 
\label{eq:colourmagcuts1}
\eea
\bea
{\rm Cut\,\,II :} & i < 18.3 + 2d_{\perp} \,\,\,, \nonumber \\
& i < 20 \,\,\,.
\label{eq:colourmagcuts}
\eea
Note we use the $r$ band Petrosian magnitude ($r_{Petro}$) 
for consistency with the SDSS LRG target selection.
We note that Cut~I is identical (except for the 
magnitude cuts in Eqs. \ref{eq:colourmagcuts1}) to the 
SDSS LRG Cut~I, while Cut~II was chosen to yield a 
population consistent with Cut~I (see below). 
This was intentionally done to maximize the overlap between 
any sample selected using these cuts, and the SDSS LRG spectroscopic sample.
The switch to the $i$ band for Cut~II also requires explanation. As is
clear from Fig.\ref{fig:lrg_spectrum}, the 4000 \AA~break is redshifting through 
the $r$ band throughout the fiducial redshift range of Cut II. This implies 
that the K-corrections to the $r$ band are very sensitive to redshift; the $i$ band
K-corrections are much less sensitive to redshift allowing for a more
robust selection.

Finally, we augment the star-galaxy separation from SDSS with the following 
cuts designed to minimize stellar contamination,
\bea
{\rm Cut\,\,I :} & r_{PSF} - r > 0.3 \,\,,\nonumber \\
{\rm Cut\,\,II :} & i_{PSF} - i > 0.2(21 -i) \,\,, \nonumber \\
& r_{deV} > 0.2 \,\,,
\eea
where $(r,i)_{PSF}$ are the SDSS PSF magnitudes, while $r_{deV}$ is 
the deVaucouleurs radius of the galaxy in arcseconds.

\subsection{Angular and Redshift Distributions}

\begin{figure}
\begin{center}
\leavevmode
\includegraphics[width=3.0in]{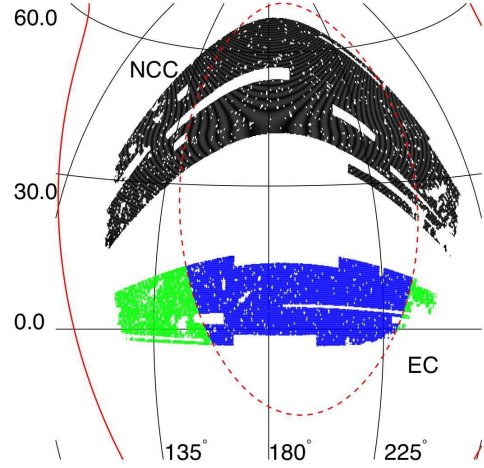}
\end{center}
\caption{The angular selection function of the LRGs with the 
``Northern Celestial Cap'' (black) and the 
``Equatorial Cap'' (blue) shown.
The lightly shaded (green) region of the Equatorial cap ($b < 45^{\circ}$, 
shown as a dashed line) is excluded because of possible stellar contamination. The 
gaps in the selection function are due to missing data and exclusion around bright 
stars. Also shown is the Galactic equator (solid line).
}
\label{fig:lrgmask}
\end{figure}

Applying the above selection criteria to the $\sim $ 5500 degrees of
photometric SDSS imaging considered in this paper yields a catalog
of approximately 900,000 galaxies. We pixelize these galaxies as a number
overdensity, $\delta_g=\delta n/\bar n$, onto a HEALPix pixelization \citep{1999elss.conf...37G}
of the sphere, with 3,145,728 pixels (HEALPix resolution 9). We exclude regions where
the extinction in the $r$-band \citep{1998ApJ...500..525S} exceeds 0.2 
magnitudes, as well as masking regions around stars in the Tycho astrometric 
catalog \citep{2000A&A...355L..27H}. We also exclude data from the three 
southern SDSS stripes due to
difficulties in photometrically calibrating them relative to the rest of the data,
due to the lack of any overlap.
The resulting angular selection function is 
shown in Fig.~\ref{fig:lrgmask}. The angular coverage naturally divides
into two regions, which we refer to as the ``Northern Celestial
Cap'' (NCC) and the ``Equatorial Cap'' (EC), based on their positions 
on the celestial sphere. As discussed below, we additionally excise
regions in the EC with $b < 45^{\circ}$ due to possible stellar contamination.
The final angular selection function covers a solid angle of 2,384 square degrees
(181,766 resolution 9 HEALPix pixels) in the NCC, and 1,144 square
degrees (87,263 resolution 9 HEALPix pixels) in the EC.

\begin{figure}
\begin{center}
\leavevmode
\includegraphics[width=3.0in]{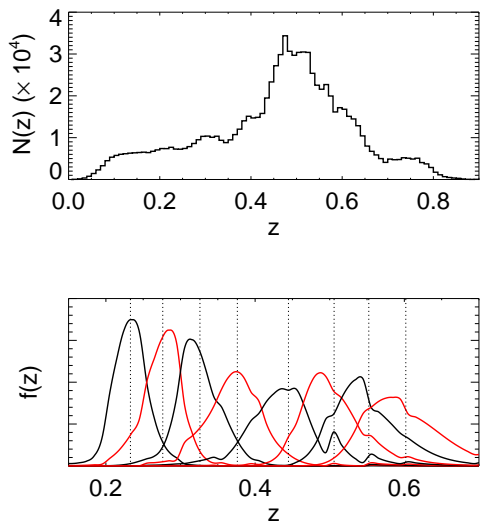}
\end{center}
\caption{(Top) The photometric redshift distribution of the 
LRG sample. (Bottom) The deconvolved selection functions for photometric redshift slices 
with $\Delta z=0.05$ from $z=0.2$ to $z=0.6$. The dotted lines are 
the mean redshifts of each of the slices. 
}
\label{fig:zslice}
\end{figure}

The calibration and accuracy of photometric redshift algorithms for this sample have been discussed
in detail by \cite{2005MNRAS.359..237P}. We compute photometric redshifts for all the galaxies
in the sample using the simple template fitting algorithm described there; these redshifts
have calibrated errors of $\sigma_{z} \sim 0.025$ at $z \sim 0.2$ that
increase to $\sigma_{z} \sim 0.05$ at $z\sim 0.6$. The resulting 
photometric redshift distribution is in Fig.~\ref{fig:zslice}. The sample is divided
into 8 photometric redshift slices of thickness $\Delta z = 0.05$ ($z00$ through
$z07$), and the underlying redshift distributions for each slice are estimated using the deconvolution 
algorithm presented in the above reference. These redshift distributions are plotted
in Fig.~\ref{fig:zslice}, while properties of the different slices are summarized in 
Table~\ref{tab:lrgdata}.

\begin{table}
\begin{tabular}{cccccc}
\hline
Label & $z_{mid}$ & $z_{mean}$ & $N_{gal}$ & $N_{gal}$ & $b_{g}$ \\
& & & (NCC) & (EC) & \\
\hline
z00 &0.225 &0.233 &   16983 &    7942 &  1.74$\pm$  0.05 \\
z01 &0.275 &0.276 &   20377 &    9283 &  1.52$\pm$  0.06 \\
z02 &0.325 &0.326 &   21759 &   10768 &  1.67$\pm$  0.07 \\
z03 &0.375 &0.376 &   28345 &   12706 &  1.94$\pm$  0.06 \\
z04 &0.425 &0.445 &   41527 &   18767 &  1.75$\pm$  0.06 \\
z05 &0.475 &0.506 &   71131 &   33000 &  1.73$\pm$  0.04 \\
z06 &0.525 &0.552 &   65324 &   30281 &  1.80$\pm$  0.04 \\
z07 &0.575 &0.602 &   46185 &   20504 &  1.85$\pm$  0.05 \\

\hline
\end{tabular}
\caption{\label{tab:lrgdata} Descriptions of the 8 $\Delta z=0.05$ redshift slices;
$z_{mid}$ is the midpoint of the redshift interval, while $z_{mean}$ is the mean redshift
of the slice. Also listed are the number of galaxies ($N_{gal}$) for 
the ``Northern Celestial Cap'' (NCC),
and the ``Equatorial Cap'' (EC), and the linear bias of each redshift slice, $b_{g}$.}
\end{table}

\subsection{Sample Systematics}

\begin{figure}
\begin{center}
\leavevmode
\includegraphics[width=3.0in]{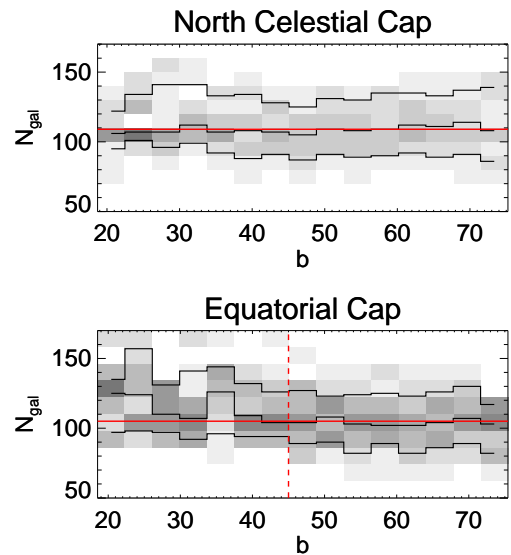}
\end{center}
\caption{The average number of LRGs per resolution 6 HEALPix pixel (approximately $1\,{\rm deg}^{2}$
in area) as a function of Galactic latitude, for the two disjoint caps. The contours are 
16\%, 50\% and 84\%. There is some evidence for stellar contamination (see text for more details) at
low Galactic latitudes in the Equatorial Cap; excising the region $b < 45^{\circ}$ removes 
the problematic regions.}
\label{fig:galdense_bval}
\end{figure}

\begin{figure}
\begin{center}
\leavevmode
\includegraphics[width=3.0in]{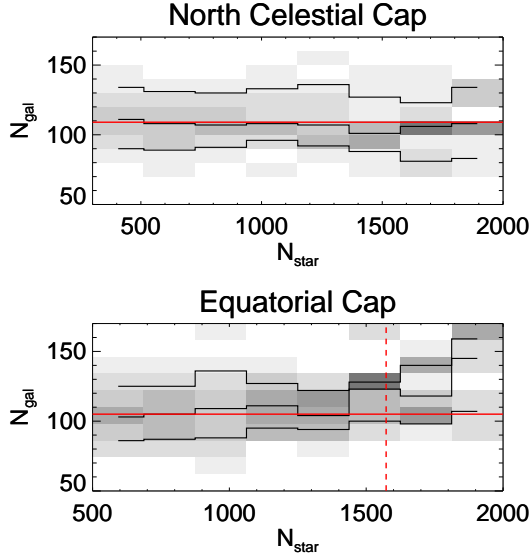}
\end{center}
\caption{The same as Fig.~\ref{fig:galdense_bval} except now as a function of stellar density.
The stellar density is estimated analogous to the galaxy density, selecting 
stars with PSF fluxes $r_{PSF}$ between 18.0 and 19.5. The vertical line
shows the position of the $b < 45^{\circ}$ cut in the Equatorial Cap.
}
\label{fig:galdense_stellar}
\end{figure}

There are a number of systematic effects in photometric samples that
contaminate clustering - stellar contamination,
angular and radial modulation of the selection due to seeing variations, extinction,
and errors in our modelling of the galaxy population. 
Fig.~\ref{fig:galdense_bval}
plots the areal LRG density as a function of Galactic latitude; one would expect any leakage
in the star-galaxy separation to increase at lower latitudes where the stellar density is higher. 
We see no increase for the NCC, but observe an increase for $b<45^{\circ}$ for the EC.
This is further borne out by Fig.~\ref{fig:galdense_stellar}, where we plot the LRG density versus
the density of stars with SDSS PSF magnitudes $18.0 < r_{PSF} < 19.5$, where the magnitude limits were
chosen so that the SDSS star-galaxy separation is essentially perfect. Although
the precise signature of such contamination on the clustering signal is unclear, we choose to be 
conservative and exclude regions below $b = 45^{\circ}$ (Figs.~\ref{fig:galdense_bval} and 
~\ref{fig:galdense_stellar}); this reduces the area of the EC by 25\%.

To understand the nature of this contamination, we 
consider the subset of galaxies for which SDSS has measured spectra.  
We find that 118,053 (13.1\%) galaxies in the photometric sample have 
measured spectra. Of these, 662 (0.56\%) are unambiguously classified as stars 
(475 objects) or quasars (187 objects). The quasars are at low 
($0.1 < z < 0.25$) redshifts, while the stars are almost entirely
K and M stars, and are preferentially at lower Galactic latitudes, consistent with
the above. Inspecting the imaging data shows that these are 
either late-type stars blended with other stars (approximately $2/3$),
late-type stars blended with background galaxies (approximately $1/3$), and 
a smattering of star-artefact blends. Note that this explains the 
dependence with Galactic latitude and stellar density; one would naively expect the number of 
star-star blends to scale as the square of the stellar density, while
the star-galaxy and star-artefact blends should roughly scale as the stellar density.
We emphasize that 
the levels of contamination obtained this way are approximate, since the spectroscopic
survey has a brighter apparent luminosity limit than our photometric catalog, and
the contamination could increase with decreasing luminosity.

\begin{figure}
\begin{center}
\leavevmode
\includegraphics[width=3.0in]{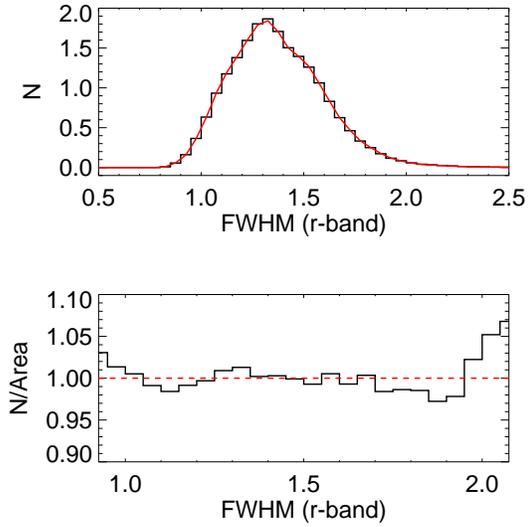}
\end{center}
\caption{(top) The histogram is the (normalized) distribution of galaxies as a function 
of the PSF FWHM (measured in arcseconds) in the $r$ band. The (red) curve is the 
fraction of the total survey area with the same PSF FWHM. The agreement
between them suggests that the galaxy selection algorithm is 
unaffected by seeing. (bottom) The galaxy surface density as a function of seeing.
The two distributions are identical at the $2\%$ level except at the edges where the 
relevant survey area is negligible.
}
\label{fig:galdense_seeing}
\end{figure}

\begin{figure}
\begin{center}
\leavevmode
\includegraphics[width=3.0in]{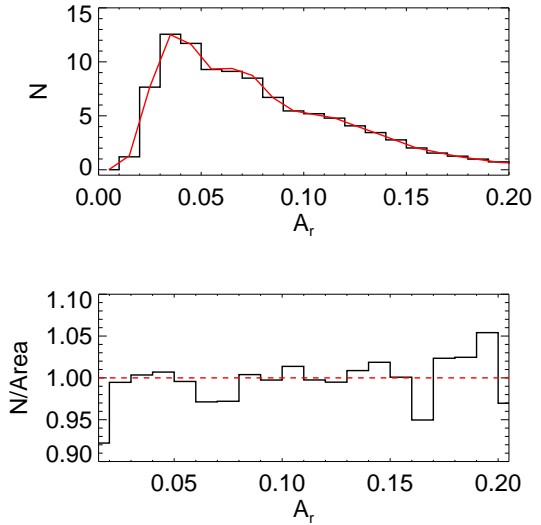}
\end{center}
\caption{Analogous to Fig.~\ref{fig:galdense_seeing} except with extinction in the $r$-band
from Schlegel et al. (\citeyear{1998ApJ...500..525S}). We truncate at $A_{r} = 0.2$ corresponding to
the cut in the angular selection function.}
\label{fig:galdense_extinct}
\end{figure}

To test for the possible modulation of the LRG selection due to angular variations
in seeing and extinction, we consider the areal density of LRGs observed
as a function of seeing (as measured by the FWHM of the $r$-band PSF) and 
extinction \citep{1998ApJ...500..525S}. These distributions are plotted in
 Figs.~\ref{fig:galdense_seeing} and 
~\ref{fig:galdense_extinct}. We find that the density is constant 
to $\sim 2\%$ over most of the range of seeing and extinction in the survey. We 
do observe deviations at the very edges of the distributions, but the total 
area with these extremes in seeing and extinction is negligible (as seen in the top
panels of the figures), and therefore, do not affect clustering measurements.

Finally, to test sample uniformity as a function of redshift, we consider the 
luminosity distribution as a function of redshift. A constant luminosity distribution
over the redshift range would suggest that we were selecting comparable populations 
of galaxies. A complication is that we must use photometric redshifts to compute
absolute magnitudes; biases in the photometric redshifts could 
alter the inferred luminosity distributions. We estimate the magnitude 
of such biases from Table~\ref{tab:lrgdata}. At low redshifts, the 
photometric redshifts are essentially unbiased, whereas at high redshifts, the 
photometric redshifts underestimate the true redshift by about $\Delta z = 0.025$,
which translates into an overestimation of the magnitude by about $\Delta M = 
0.1 - 0.15$ magnitudes.

\begin{figure}
\begin{center}
\leavevmode
\includegraphics[width=3.0in]{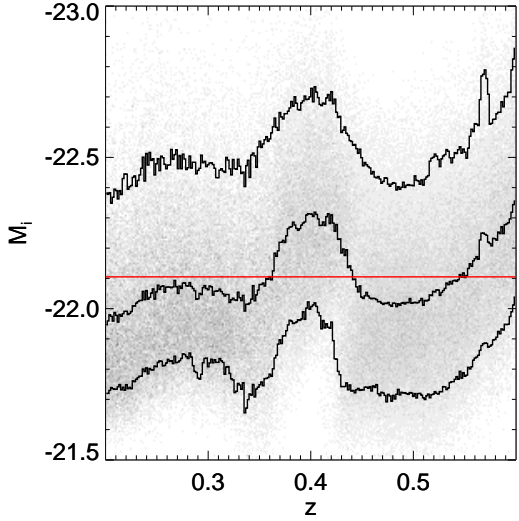}
\end{center}
\caption{The conditional $i-$band magnitude distribution as a function of 
redshift. The absolute magnitude is computed assuming the photometric redshift.
The contours show the 16\%, 50\%, and 84\% levels, while the horizontal (red)
line is the median magnitude for the entire sample. The glitch at $z \sim 0.4$
corresponds to the transition between Cut~I and Cut~II LRGs, while the 
increase at $z \sim 0.55$ is due to the magnitude limit in Cut~II.}
\label{fig:magzdist}
\end{figure}

The observed conditional luminosity distribution as a function of redshift is
in Fig.~\ref{fig:magzdist}. The median luminosity is constant to approximately
$\Delta M = 0.1$ over the redshift range of interest, with a width of $\sim 0.7$
magnitudes (compared with a potential bias of $0.15$ magnitudes above). The 
distribution has two distinguishing features, a glitch at 
$z \sim 0.4$ and increasing luminosities at higher redshifts. The 
glitch at $z \sim 0.4$ corresponds to the transition between Cut~I and Cut~II 
at the point where colour tracks bend sharply in Fig.~\ref{fig:lrg_colordiag},
and highlights a difficulty in uniformly selecting galaxies in that region. The 
increase in luminosities at $\sim z=0.55$ is due to the magnitude limits
imposed in Cut~II. Except for these two features, we conclude that our 
selection criteria yield an approximately uniform galaxy population
from  $z=0.2$ to $z=0.55$.

\section{The Angular Power Spectrum}
\label{sec:angular}

\subsection{Projections on the sky}
\label{sec:theory1}

We relate the projected angular power spectrum to
the underlying three dimensional power spectrum; our derivation 
follows the discussion in 
\cite{2001ApJ...555..547H} \citep[see also][and references therein]{2002ApJ...571..191T}.
We describe the galaxy distribution by an 
isotropic 3D density field, $\delta_{g,3D}$, and its power spectrum $P(k)$
defined by,
\be
\langle \delta_{g,3D}(\kvec) \delta_{g,3D}^{*}(\kvec') \rangle 
= (2\pi)^{3} \delta^{3}(\kvec - \kvec') P_{g}(k) \,\,.
\label{eq:3dpower}
\ee
Projecting this density field on the sky along $\nhat$, we obtain,
\be
\delta_{g}(\nhat) = \frac{1}{\int \,dy\,\phi(y)} \int \, dy\, \phi(y) \delta_{g,3D}(y, y\nhat) \,\,,
\label{eq:deltag}
\ee
where $y$ is the comoving distance, and $\phi(y)$ is the radial selection function. For now,
we ignore the effect of peculiar velocities, and therefore do not distinguish between
real and redshift space quantities. Fourier transforming the 3D density field and making
use of the identity,
\be
e^{-i \kvec \cdot \nhat y} = \sum_{l=0}^{\infty} (2l+1) i^{l} j_{l}(ky) P_{l}(\khat \cdot \nhat) \,\,,
\ee
we obtain,
\bea
& & \delta_{g}(\nhat) = \int \, dy \, f(y) \int\frac{d^{3}k}{(2\pi)^3} \delta_{g,3D}(y, \kvec) \nonumber  \\
& & \times \sum_{l=0}^{\infty} i^{l} (2l+1) j_{l}(ky) P_{l}(\nhat \cdot \khat)  \,\,,
\label{eq:deltagl1}
\eea
where $j_{l}(x)$ and $P_{l}(x)$ are the $l^{\rm th}$ order spherical Bessel functions and 
Legendre polynomials respectively. We define the weighting function, $f(y)$ by 
\be
f(y) \equiv \frac{\phi(y)}{\int \,dy\,\phi(y)} \,\,.
\ee
Since the density field is isotropic, we expand it in Legendre polynomials to obtain,
\be
\delta_{g,l} = i^{l} \int dy \, f(y) \int\frac{d^{3}k}{(2\pi)^3} \delta_{g,3D}(y, \kvec) 
j_{l}(ky) \,\,.
\ee

\begin{figure}
\begin{center}
\leavevmode
\includegraphics[width=3.0in]{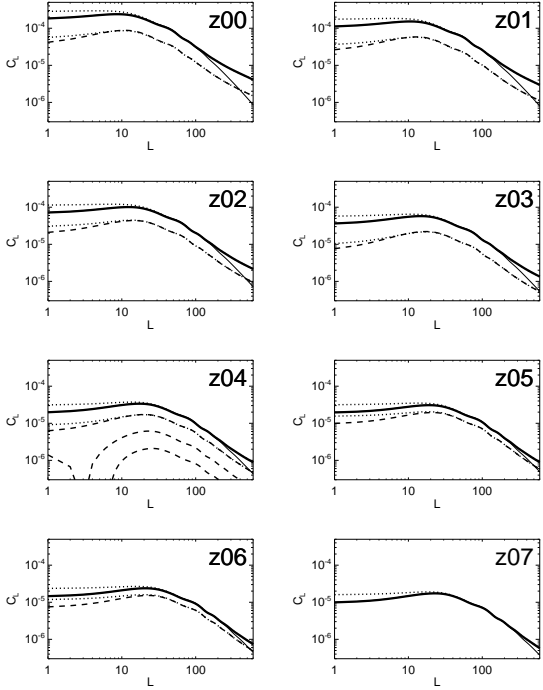}
\end{center}
\caption{The theoretical angular power spectra for each of the redshift slices 
in Fig.~\ref{fig:zslice}. The heavy and light solid lines show nonlinear and linear auto power
spectra, while the dashed lines show the cross power spectra with the 
adjacent slice at higher redshift. The dotted lines show the effect of redshift space distortions
on both the auto and cross power spectra, assuming $\beta = 0.3$. The panel $z04$ also
shows the cross correlation power with $z06$ and $z07$.
}
\label{fig:cltheory}
\end{figure}

In order to proceed, we assume that the selection function is narrow in redshift, allowing
us to ignore the evolution of the density field. The above equation can then be written 
as
\be
\delta_{g,l} = i^{l} \int\frac{d^{3}k}{(2\pi)^3} \delta_{g,3D}(\kvec) W_{l}(k) \,\,,
\label{eq:deltagl2}
\ee
where we implicitly assume that the density field is at the median redshift of
the selection function. The window function, $W_{l}(k)$, describes the mapping of $k$
to $l$ and is given by, 
\be
W_{l}(k) = \int dy \, f(y) j_{l}(ky) \,\,.
\ee
It is now straightforward to 
compute the angular power spectrum,
\be
C_{l} \equiv \langle \delta_{g,l} \delta_{g,l}^{*} \rangle = 4 \pi \int
\frac{dk}{k} \Delta^{2}(k) W_{l}^{2}(k) \,\,,
\label{eq:cl}
\ee
where $\Delta^{2}(k)$ is the variance per logarithmic wavenumber, 
\be
\Delta^{2}(k) \equiv \frac{1}{(2\pi)^3} 4\pi k^{3} P(k) \,\,.
\ee
Similarly, the cross correlation between two selection functions, $\phi_{1}$ and $\phi_{2}$, is
give by
\be
C^{12}_{l} = 4 \pi \int
\frac{dk}{k} \Delta^{2}(k) W_{l,1}(k) W_{l,2}(k)\,\,.
\ee

We have not distinguished between the galaxy and matter power spectrum 
above. On large scales, we simply assume
\be
P_{g}(k) = b_{g}^{2} P(k) \,\,,
\ee
where $P_{g}(k)$ and $P(k)$ are the galaxy and matter power spectra respectively, and 
$b_{g}$ is the linear galaxy bias. This
is a good approximation on large scales \citep{1998ApJ...504..607S}, 
but breaks down on smaller
scales; we defer a discussion of its regime of validity, as well as the nonlinear 
evolution of the power spectrum to a later section.

Fig.~\ref{fig:cltheory} shows the predicted angular power spectra for the
eight redshift distributions in Fig.~\ref{fig:zslice} assuming our fiducial
cosmology; also shown are the cross-correlation 
power spectra for adjacent slices. We assume $b_{g}=1$, and use the
\texttt{halofit} prescription \citep{2003MNRAS.341.1311S} to evolve the matter power
spectrum into the nonlinear regime. The increase in the amplitude of the 
power spectrum on large scales (low $L$) with decreasing redshift 
is due to the linear growth of structure, while 
the increase in power on small scales (large $L$)
is from the nonlinear collapse of structures. The ``knee'' in the power
spectrum between $L \sim 10 - 30$ corresponds to the turnover
in the 3D power spectrum $P(k)$, where the shape changes from 
$P(k) \sim k$ to $P(k) \sim k^{-3}$ (in the linear regime). This scale
corresponds to the horizon at matter-radiation equality and 
is constant in comoving coordinates. However, with increasing radial 
distances to the redshift slices, the apparent angular size decreases with
redshift, and we see the ``knee'' shift from 
low $L$ (large angular scales) at low redshifts 
to high $L$ (small angular scales) at high redshifts. This illustrates
the potential use of the power spectrum as a standard ruler for cosmography;
given the size of the horizon at matter-radiation equality (independent of 
dark energy), one can probe the evolution of the universe during the 
dark energy dominated phase. A second such standard ruler is the 
baryonic oscillations in the matter power spectrum visible at
$L\sim 100$. However, its amplitude is suppressed in the individual angular
power spectra by the smoothing due to the thickness of the redshift slices.

Finally, we note that the cross-correlation between adjacent slices is non-negligible.
This is easily understood by considering Fig.~\ref{fig:zslice}, where we note that there
is considerable overlap between adjacent slices. Furthermore, this overlap 
increases with increasing redshift due to larger photometric redshift errors; this 
too is reflected in the cross-correlations. Going to more widely separated slices 
reduces the cross-correlation due to smaller overlaps. Note that the 
level of correlation seen in Fig.~\ref{fig:cltheory} is only true on large
scales; on smaller scales, uncorrelated Poisson noise (since the galaxy
samples are disjoint) erases these correlations.

\subsubsection{Redshift Space Distortions}
\label{sec:theory2}

The above discussion ignored the effect of peculiar velocities on the 
observed clustering power spectrum. For broad redshift selection functions, 
the projection on to the sphere erases redshift space distortions; 
however, as the selection function becomes narrow, they become more
important. We calculate their effect below, following the formalism of 
\cite{1994MNRAS.266..219F}.

We start with Eq.~\ref{eq:deltag},
\be
1+\delta_{g}(\nhat) = \int \, dy\, f(s) [1+\delta_{g,3D}(y, y\nhat)] \,\,,
\label{eq:deltag_red}
\ee
where we have now written the weighting function as a function of redshift 
distance, $s = y + {\mathbf v}\cdot\nhat$, and we have left the monopole 
contribution to the projected galaxy density explicit. Assuming the
peculiar velocities are small compared with the thickness of the redshift
slice, we Taylor expand the weight function to linear order,
\be
f(s) \approx f(y) + \frac{df}{dy} ({\mathbf v}(y\nhat)\cdot\nhat) \,\,.
\ee
Substituting this expression into Eq.~\ref{eq:deltag_red}, we note that at linear order,
redshift space distortions only imprint fluctuations on the monopole component
of the galaxy density. This allows us to separate the 2D galaxy density
into two terms, $\delta_{g} = \delta_{g}^{0} + \delta_{g}^{r}$, where $\delta_{g}^{0}$
is the term discussed above, while $\delta_{g}^{r}$ are the redshift space 
distortions. Fourier transforming the velocity field, we find that,
\be
\delta_{g}^{r}(\nhat) = \int \, dy\, \frac{df}{dy} \int \frac{d^{3} k}{(2 \pi)^{3}} 
{\mathbf v}(\kvec) \cdot \nhat e^{-i \kvec \cdot \nhat y} \,\,.
\label{eq:deltag_v}
\ee
The linearized continuity equation allows us to relate the velocity and density
perturbations, 
\be
{\mathbf v}(\kvec) = -i \beta \delta_{g}(\kvec) \frac{\kvec}{k^{2}}\,\,,
\label{eq:velocity}
\ee
where $\beta$ is the redshift distortion parameter given approximately by
$\beta \approx \Omega_{m}^{0.6}/b_{g}$.
Substituting Eq.~\ref{eq:velocity} into Eq.~\ref{eq:deltag_v} 
and taking the Legendre transform, we can rewrite this equation
in the form of Eq.~\ref{eq:deltagl2} where the window function now has an 
additional component given by,
\be
W^{r}_{l}(k) = \frac{\beta}{k} \int \, dy \, \frac{df}{dy} j_{l}'(ky) \,\,,
\ee
where $j_{l}'$ is the derivative of the spherical Bessel function with respect to 
its argument. By a repeated application of the recurrence
relation $l j_{l-1} - (l+1) j_{l+1} = (2l+1) j_{l}'$, and integrating by parts,
\bea
W^{r}_{l}(k) = \beta \int \, dy \, f(y) \left[ \frac{(2l^{2} + 2l -1)}{(2l+3)(2l-1)} j_{l}(ky)
\right. \nonumber \\
\left. - \frac{l(l-1)}{(2l-1)(2l+1)} j_{l-2}(ky) - 
\frac{(l+1)(l+2)}{(2l+1)(2l+3)} j_{l+2}(ky) \right]\,\,.
\eea
It is interesting to note that this result could have been equivalently derived
by starting from the \cite{1987MNRAS.227....1K} enhancement 
of the 3D power spectrum due
to redshift space distortions, $P_{g}(\kvec) \rightarrow P_{g}(\kvec)[ 1 + 
\beta (\kvec \cdot \nhat)]$, and integrating along the line of sight as in 
Sec.~\ref{sec:theory1}; the $l \pm 2$ spherical Bessel functions result
from the coupling of the $\kvec \cdot \nhat$ angular dependence to the 
Legendre polynomials. Also interesting is the $l \gg 0$ limit
of the above equation; for $l$ sufficiently large, $\int dy\,f(y) j_{l}(y)
\approx \int dy\,f(y) j_{l\pm 2}(y)$, and $W^{r}(k)$ vanishes. Physically,
this corresponds to the radial velocity perturbations being erased by the 
projection on to the sky.

Fig.~\ref{fig:cltheory} shows the effects of redshift space distortions on the 
angular power spectra for the eight redshift slices we are considering. Note 
that they contribute significantly only on the largest scales ($l < \sim 30$),
justifying our use of linear theory.

\subsection{Power Spectrum Estimation}

The theory behind optimal power spectrum estimation is now well established,
and so we limit ourselves to details specific to this discussion, and refer
the reader to the numerous references on the subject 
\citep[][and references therein]{1998ApJ...499..555T,
1998ApJ...506...64S, 2001ApJ...550...52P}.

We start by parametrizing the power spectrum with twenty step functions in 
$l$, $\tilde{C}^{i}_{l}$, 
\be
C_{l} = \sum_{i} p_{i} \tilde{C}^{i}_{l} \,\,,
\ee
where the $p_{i}$ are the parameters that determine the power spectrum. We
form quadratic combinations of the data,
\be
q_{i} = \frac{1}{2} {\mathbf x}^{T} {\mathbf C}_{i} {\mathbf C}^{-1}
 {\mathbf C}_{i} {\mathbf x}\,\,,
\ee
where ${\mathbf x}$ is a vector of pixelized galaxy overdensities, 
${\mathbf C}$ is the covariance matrix of the data, and ${\mathbf C}_{i}$
is the derivative of the covariance matrix with respect to $p_{i}$.
The covariance matrix requires a prior power spectrum to account for
cosmic variance; we estimate the prior by computing an estimate of the
power spectrum with a flat prior and then iterating once.
We also construct the Fisher matrix,
\be
F_{ij} = \frac{1}{2} 
  {\rm tr} \left[{\mathbf C}_{i} {\mathbf C}^{-1} {\mathbf C}_{j} 
{\mathbf C}^{-1}\right] \,\,.
\ee
The power spectrum can then be estimated, $\hat{\mathbf p} =
{\mathbf F}^{-1} {\mathbf q}$, with covariance matrix ${\mathbf F}^{-1}$.

A final note on implementation - the dimension of the data covariance matrix
is given by the number of pixels in the data. This quickly makes 
any direct implementation of this algorithm impractical. We therefore use
the algorithm outlined by \cite{2003NewA....8..581P}, modified 
for a spherical geometry as in \cite{2004PhRvD..70j3501H}.

\subsection{Simulations}

\begin{figure}
\begin{center}
\leavevmode
\includegraphics[width=3.0in]{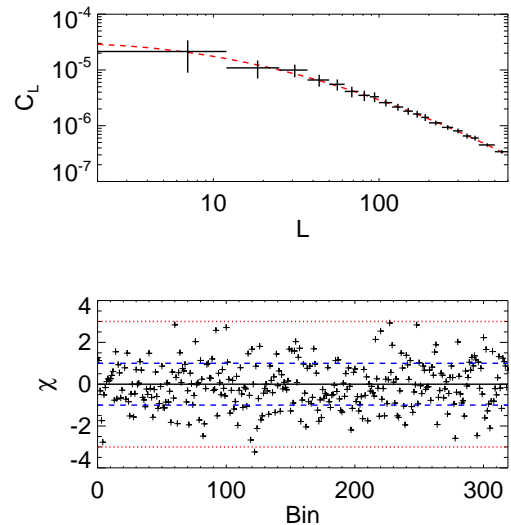}
\end{center}
\caption{ (top) The average recovered power spectrum from 100 simulated realizations. 
The dashed (red) line is the input power spectrum. The error bars are 
the errors per realization, and are not the error on the mean. Note that we
have suppressed the input power spectrum by a constant factor relative to the
expected power to avoid getting $\delta > 1$; the galaxy number density was
boosted by the same factor to reduce the shot noise.(bottom) $\chi$ comparing
the errors on the power spectrum derived from the Fisher matrix versus
those obtained from the run to run variance of the simulations for 
each of the eight redshift slices and the two angular caps. Assuming the density field is
Gaussian, the error on the power spectrum errors ($\sigma$) is 
$\sigma/\sqrt{2N}$, where $N$ is the number of the simulations. Also shown
are the $1\sigma$ and $3\sigma$ lines.
}
\label{fig:plotsims}
\end{figure}

Before applying the above algorithm to the LRG catalog, we apply it to simulated
data. In addition to testing the accuracy of our power 
spectrum code, we would also like to understand the correlations between the 
NCC and the EC, allowing us to combine separate power spectrum measurements. 

In order to do so, we use the prior power spectra 
for each redshift slice to simulate 
a Gaussian random field over the entire sphere. We then Poisson distribute galaxies 
with probability $(1+\delta)/2$ over the survey region, trimmed with the angular selection
function. One technical complication \citep{2001ApJ...550...52P} is that
the measured amplitude of the power spectrum results in a number of points with $|\delta| > 1$, making
simple Poisson sampling impossible. To avoid this, we suppress the 
power spectrum by a constant factor, and boost the number density of galaxies by 
the same factor to ensure that the shot noise is similarly suppressed. 
We generate 100 such simulations for the eight redshift slices and both angular caps 
separately, matching the observed numbers of galaxies in each case; although the different redshift
bins are uncorrelated, the angular caps are based on correlated density fields. 
This allows us to estimate the covariance between power spectrum measurements made
for the different caps. Our goal here is not to realistically 
simulate galaxy formation, but to test our pipelines, and the resulting measurements and 
errors; Gaussian simulations are sufficient for this purpose.

The results from one set of 100 simulations are shown in the top panel of Fig.~\ref{fig:plotsims};
the recovered power spectrum agrees well with the input power spectrum.
The bottom panel of the same plot compares the errors as measured by the inverse of the
Fisher matrix with those obtained from the run to run variance of the simulations.
Assuming Gaussianity, these errors should themselves have a relative error given 
by $\Delta \sigma/\sigma = 1/\sqrt{2 N}$ where $N=100$ is the number of the simulations.
As evident from the figure, the run to run variance agrees (within the expected
errors) with the errors from the Fisher matrix.
 
\subsection{The Angular Power Spectrum}
\label{sec:angpower}

\begin{figure*}
\begin{center}
\leavevmode
\includegraphics[width=6.0in]{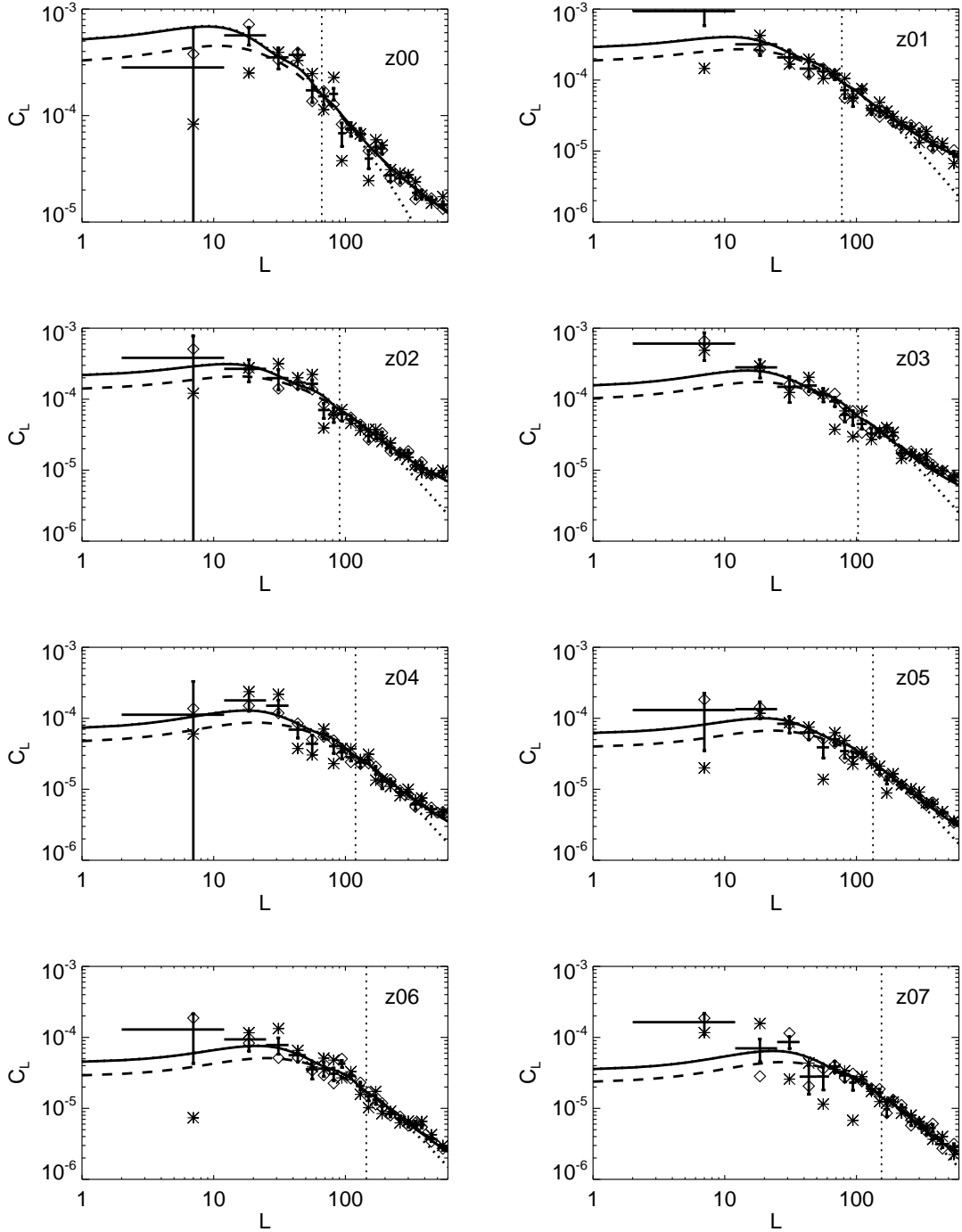}
\end{center}
\caption{The measured angular power spectrum for the 8 redshift bins. The crosses
show the power spectrum (and measured errors) of Cap I and Cap II combined, while the 
diamonds and stars are the measured power spectra of Cap 1 and Cap II separately. The 
solid lines are the predicted nonlinear power spectra for our fiducial cosmological
model, while the dotted line shows the linear prediction.
The dashed line is the nonlinear power spectrum for a model with negligible baryonic
content. The vertical line marks the nominal nonlinear scale given by $k=0.1 h \rm{Mpc}^{-1}$.
}
\label{fig:cl2d}
\end{figure*}

Fig.~\ref{fig:cl2d} shows the measured angular power spectrum for the eight redshift slices,
with the two angular caps being measured separately. The difficulty with processing 
the two angular caps simultaneously is that errors in photometric calibration 
masquerade as large scale power. While it is possible to control these systematics in
regions with overlaps in the data, the two angular caps are disconnected; therefore, any
relative calibration between the two caps must be indirect (eg. considering data
taken on the same night, and assuming that the calibration is constant through 
the night). Unfortunately, the expected power on these scales is also small ($\Delta^{2}
\sim 10^{-3}$), and so we choose to be conservative and measure the angular power spectrum
for the caps separately. We combine these using the simulations 
of the previous section to correctly take the covariances between the two caps into 
account. In order to avoid mixing power between different angular scales, we 
simply use constant weights 
proportional to the area (0.67 and 0.33 for the NCC and EC respectively); these are
approximately the same weights that one would have obtained by inverse variance weighting.
The final results are in Fig.~\ref{fig:cl2d}. 

\subsection{Bias and $\beta$}

\begin{figure}
\begin{center}
\leavevmode
\includegraphics[width=3.0in]{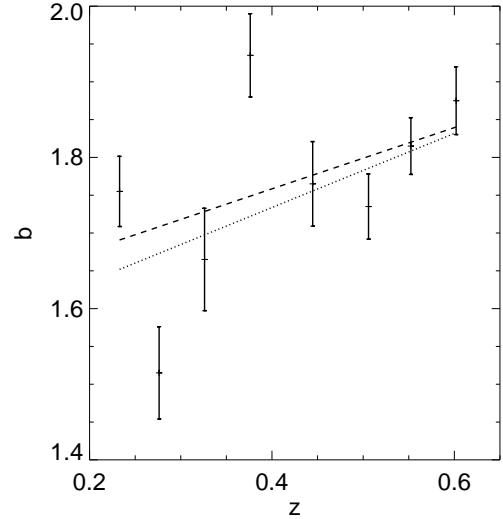}
\end{center}
\caption{Bias as function of redshift, as estimated for the 
eight redshift slices, marginalizing over redshift space 
distortions. Note that the fourth slice, with its
anomalous bias, corresponds in redshift to the glitch seen in
Fig.~\ref{fig:magzdist}. The dashed line shows the best linear fit
to the all eight bias values, while the dotted line 
excludes the fourth data point. Note that we have ignored the correlations
between the different redshift slices for the fit.
}
\label{fig:lrg_bias}
\end{figure}

An immediate question is whether the power spectra in Fig.~\ref{fig:cl2d}
are consistent with being derived from a single 3D power spectrum, appropriately 
normalized to account for bias and the evolution of structure. We start with the linear
3D power spectrum for our fiducial cosmology, and project it to a 2D power spectrum
$C_{l,gg}$, using the formalism of Sec.~\ref{sec:theory1}. We also compute the effect
of redshift space distortions, whose normalization we parametrize by $\beta$, along the
lines of Sec.~\ref{sec:theory2}, giving us two more power spectra, $C_{l,gv}$ and $C_{l,vv}$.
The total power spectrum is,
\be
C_{l} = b_{g}^{2} \left(C_{l,gg} + 2\beta C_{l,gv} + \beta^{2} C_{l,vv} \right)\,\,,
\ee
where $b_{g}$ is the linear bias of the LRGs. The three power spectra represent 
correlations of the galaxy density with itself ($gg$), the velocity
perturbations (the source of linear redshift distortions) with itself ($vv$),
and their cross correlation ($gv$). We also note (as emphasized in Sec.~\ref{sec:theory2})
that the redshift distortions only affect the largest scales; therefore, 
the linear assumption is justified. We can now explore the $\chi^{2}$ likelihood surface
as a function of $b$ and $\beta$ for each of the redshift slices. In practice, $\beta$ 
is not strongly constrained by these data, and so we marginalize over it when estimating
the bias.

The best fit models are compared with the data in Fig.~\ref{fig:cl2d}, while the 
bias for the eight redshift slices is in Fig.~\ref{fig:lrg_bias}.
We do not fit to the entire power spectrum, but limit ourselves to 
scales larger than the nominal nonlinear cutoff at $k=0.1 h {\rm Mpc}^{-1}$; the angular scales 
corresponding to this restriction are marked in Fig.~\ref{fig:cl2d}.
Our starting hypothesis - that the angular power spectra are
derived from a single 3D power spectrum - appears to be well motivated. Interestingly, 
the \texttt{halofit} nonlinear prescription for the matter power spectrum 
fits the galaxy power spectrum data down to small scales as well. The minimum $\chi^{2}$ value is 81.6
for 62 degrees of freedom, corresponding to a probability of $4.8\%$.

Fig.~\ref{fig:lrg_bias} shows that the bias increases with increasing redshift,
as one would expect for an old population of galaxies that formed early in the first 
(and therefore most biased) overdensities. A notable exception to this trend appears
to be redshift slice $z03$; however, this redshift slice corresponds to the region of the
glitch in the luminosity-redshift distribution plotted in Fig.~\ref{fig:magzdist}. If the 
median luminosity in this redshift slice is higher than the other slices, one would 
expect a higher linear bias, consistent with what is observed.

\begin{figure}
\begin{center}
\leavevmode
\includegraphics[width=3.0in]{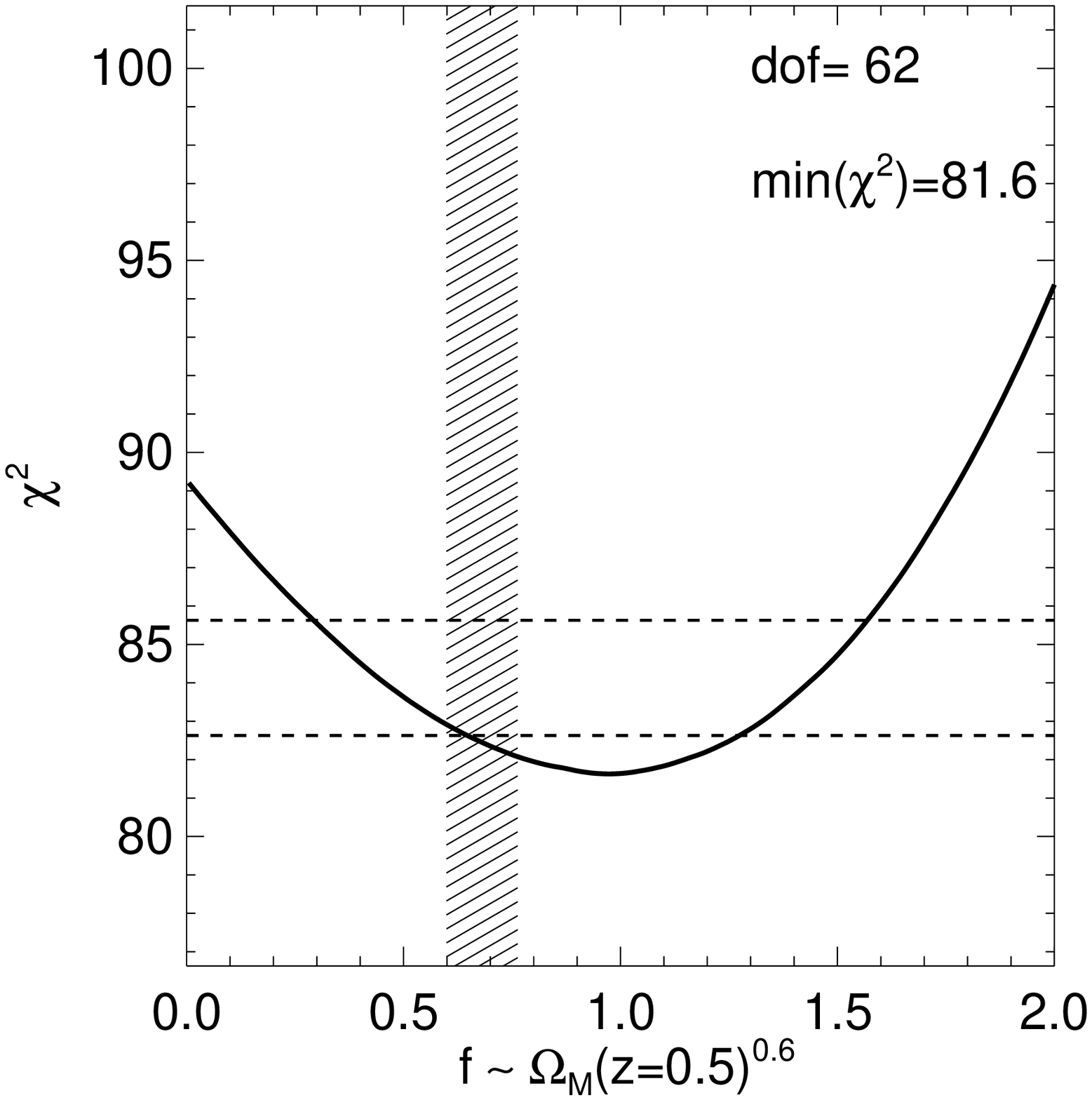}
\end{center}
\caption{$\chi^{2}$ as a function of the dimensionless
growth factor, $f = \beta b \sim \Omega_{m}^{0.6}$, marginalizing
over bias, for all eight redshift slices combined. The dashed
lines mark the $1-$ and $2-\sigma$ intervals, while the shaded 
region corresponds to the value of $\Omega_{m}^{0.6}$ between
$z=0.2$ and $z=0.6$, assuming a present day value of $0.3$. The 
$\chi^{2}$ value of 81.6, for 62 degrees of freedom has a probability
of $4.8\%$. Note that we have ignored the correlations
between the different redshift slices.
}
\label{fig:beta}
\end{figure}

In order to constrain $\beta$, we start from the definition that $\beta
\equiv f(\Omega_{M},\Omega_{\Lambda})/b$, where $f \sim \Omega_{m}(z)^{0.6}$
is the dimensionless growth factor at redshift $z$. Assuming that the error on 
$f$ is larger than the variation of $\Omega_{m}$ with redshift, we 
approximate $f$ as a constant over the depth of the survey. We can then attempt
to constrain $f$ by combining all eight redshift slices; note that for 
simplicity, we ignore the correlations between the slices and treat them as independant.
The results are in Fig.~\ref{fig:beta}. We start by noting that the 
width of the $\chi^{2}$ distribution is significantly larger than the variation 
in $f$ with redshift, justifying our starting assumption. This is 
a direct, albeit crude, measure of $\Omega_{m}(z \sim 0.5)  \approx 0.97 \pm 0.53$,
consistent with our fiducial model of $\Omega_{m}(z =0.5)=0.59$.

\subsection{Redshift correlations}

An important test of systematics is the cross correlation between different redshift
slices. For well separated slices, the cosmological correlation goes to zero on all
but the largest scales; the detection of a correlation would imply 
the presence of systematic spatial fluctuations caused by eg. stellar contamination,
photometric calibration errors, incorrect extinction corrections etc. On the other 
hand, the cosmological cross correlation is nonzero for adjacent slices due 
to overlaps in the redshift distribution, but is completely determined
theoretically by the observed auto-correlation power spectra and the input
redshift distributions. These cross correlations therefore test the accuracy 
of the estimated redshift distributions, and in particular, the wings of these
distributions where they overlap.

\begin{figure*}
\begin{center}
\leavevmode
\includegraphics[width=6.0in]{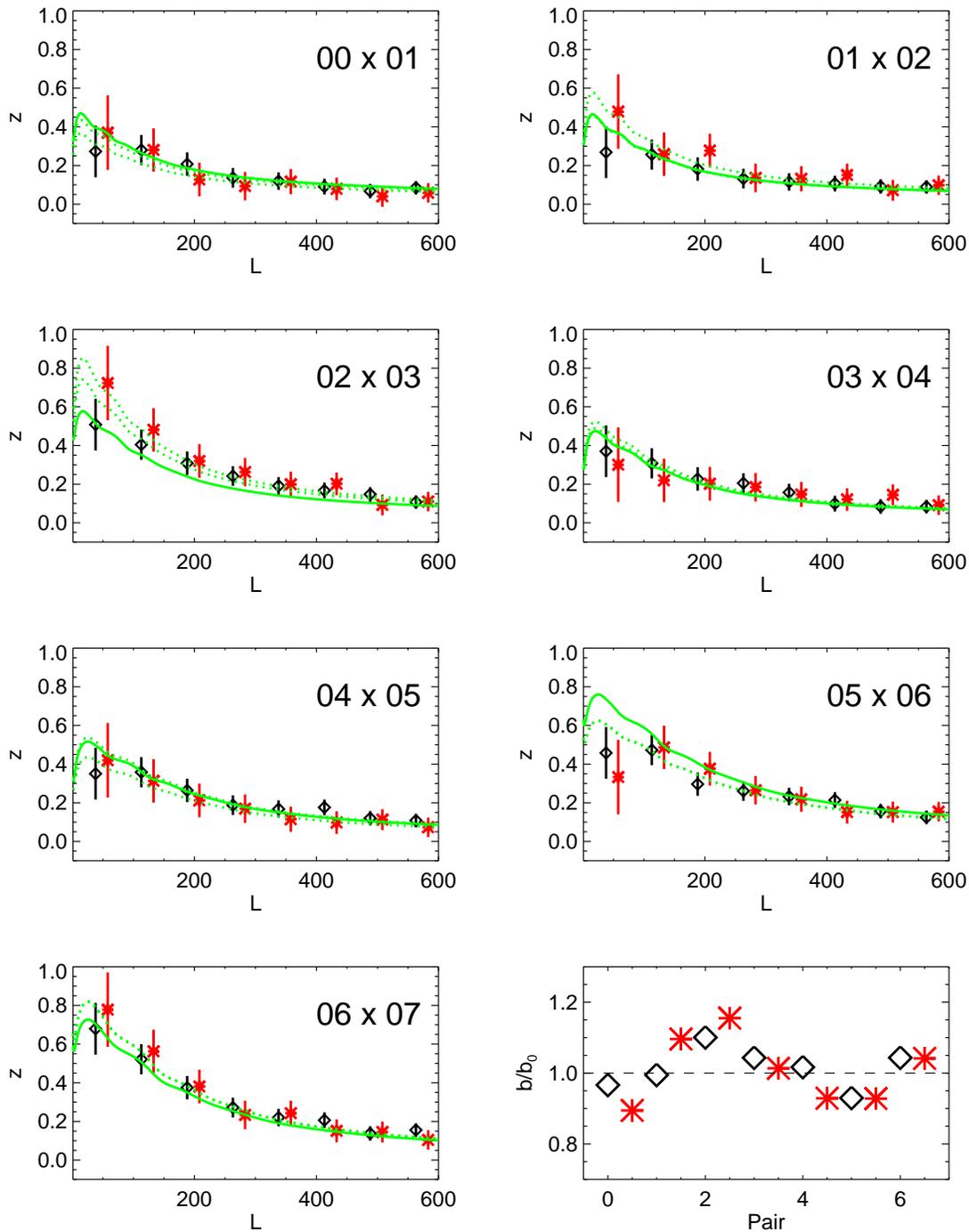}
\end{center}
\caption{The Fisher z-transform of the correlation coefficient between 
adjacent redshift slices.
The diamonds and stars are the results for the NCC and EC respectively; 
the errors are $2 \sigma$ errors.
Note that the window function is only
approximately corrected; nearby bins are therefore correlated.
The solid curve is the prediction for the cross correlation with
the bias of both the slices fixed to the autocorrelation value, while 
the dotted lines show the fits allowing a variable bias. The plot
in the lower right corner shows the best fit bias compared to the
prediction from the autocorrelation.
}
\label{fig:xcorr1}
\end{figure*}

For computational convenience, we estimate the cross correlations with a simple
pseudo-$C_{l}$ estimator,
\be
\hat{C}^{12}_{l} = \frac{1}{2l+1} \sum_{m=-l}^{l} a_{1,lm} a_{2,lm}^{*} \,\,,
\ee
where $a_{1,2,lm}$ are the spherical tranforms of the galaxy density. The 
pseudo-$C_{l}$ power spectrum is the true power spectrum convolved with
the angular mask of the survey; it is therefore convenient to work with 
the cross correlation coefficient,
\be
r^{12}_{l} \equiv \frac{\hat{C}^{12}_{l}}{\sqrt{\hat{C}^{11}_{l} \hat{C}^{22}_{l}}}
= \frac{M \star C^{12}_{l}}{\sqrt{(M \star C^{11}_{l}) (M \star C^{22}_{l})}} \,\,,
\ee
where $M \star$ represents convolutions by the angular mask. The advantage of the
cross correlation is that on scales smaller than the angular mask, the effect of the
angular mask approximately cancels, allowing for easy comparison with theory. 
It is useful to apply Fisher's 
$z$-transform \citep{1977ats..book.....K, 1992nrfa.book.....P}
\be
z = \frac{1}{2} \log\left( \frac{1+r}{1-r} \right) \,\,,
\ee
which is well described (for $l \approxgt 50$) by a Gaussian with mean,
\be
\langle z \rangle = \frac{1}{2} \log\left( \frac{1+r_{true}}{1-r_{true}} \right) + 
 \frac{r_{true}}{2(N-1)} \,\,,
\ee
and standard deviation,
\be
\sigma(z) \approx \frac{1}{\sqrt{N-3}} \,\,,
\ee
where $N \approx (2l+1)f_{sky}$ is the number of independent modes, and $r_{true}$ is the
predicted cross correlation coefficient.

\begin{figure*}
\begin{center}
\leavevmode
\includegraphics[width=6.0in]{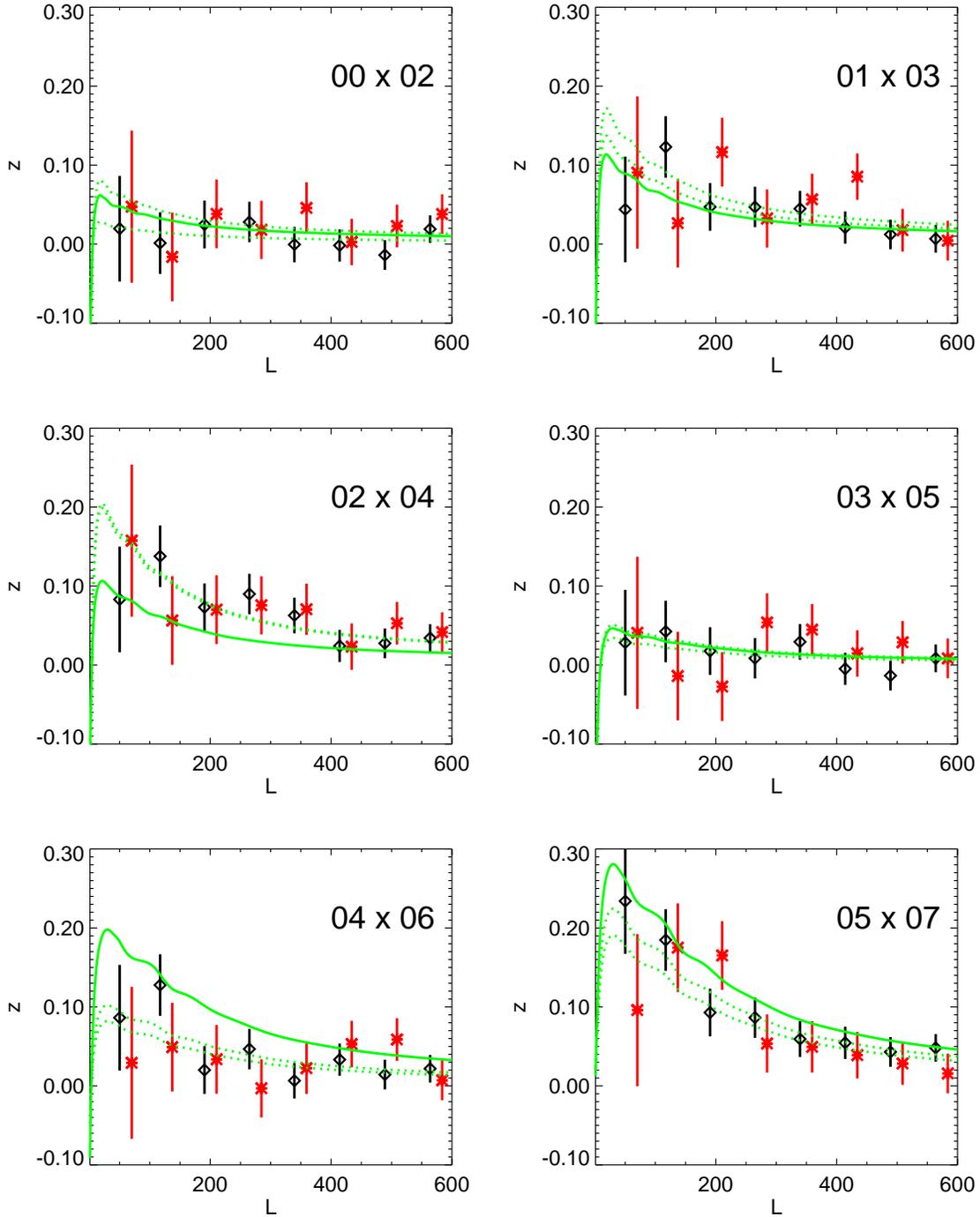}
\end{center}
\caption{The same as Fig.~\ref{fig:xcorr1} but for redshift slices separated by
one redshift bin. The overlaps at low redshifts are negligible, but increase at 
higher redshift.
}
\label{fig:xcorr3}
\end{figure*}

\begin{figure}
\begin{center}
\leavevmode
\includegraphics[width=3.0in]{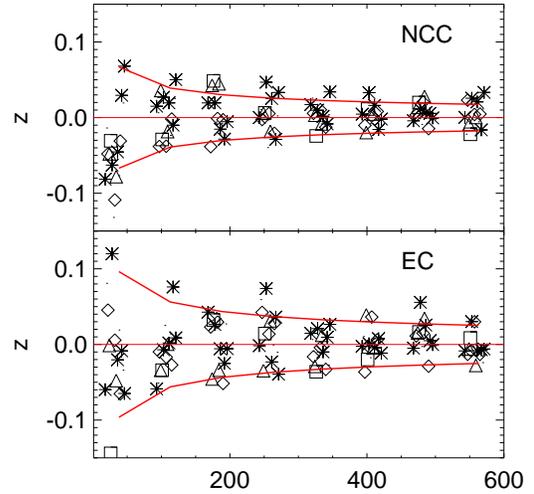}
\end{center}
\caption{The Fisher z-transform of the correlation coefficient between 
redshift slices separated by at least two redshift slices. 
The curves show the $1-$ and $3-\sigma$ contours, given the
null hypothesis of no correlations. Note that the window function is only
approximately corrected; nearby multipoles are therefore correlated.
}
\label{fig:xcorr2}
\end{figure}

Figs.~\ref{fig:xcorr1}, ~\ref{fig:xcorr3}, ~\ref{fig:xcorr2} 
show the measured cross correlations
between adjacent and more widely separated slices respectively. The absence of 
correlations between widely separated slices indicates a lack of small scale
systematics common to the different redshift slices. The cross-correlations between 
adjacent slices broadly agree with the predictions from the auto-correlations,
although there are differences at the $\sim 10\%$ level as seen in the plot in the
lower right. There are two possibilities for this disagreement. The first is that
variations in the galaxy population over a redshift slice could cause the bias in the
overlap region to differ from the value averaged over the entire slice. 
Comparing with Fig.~\ref{fig:lrg_bias}, we note that slice to slice bias variations  
of $\sim 10\%$ are consistent with the data.

\begin{figure}
\begin{center}
\leavevmode
\includegraphics[width=3.0in]{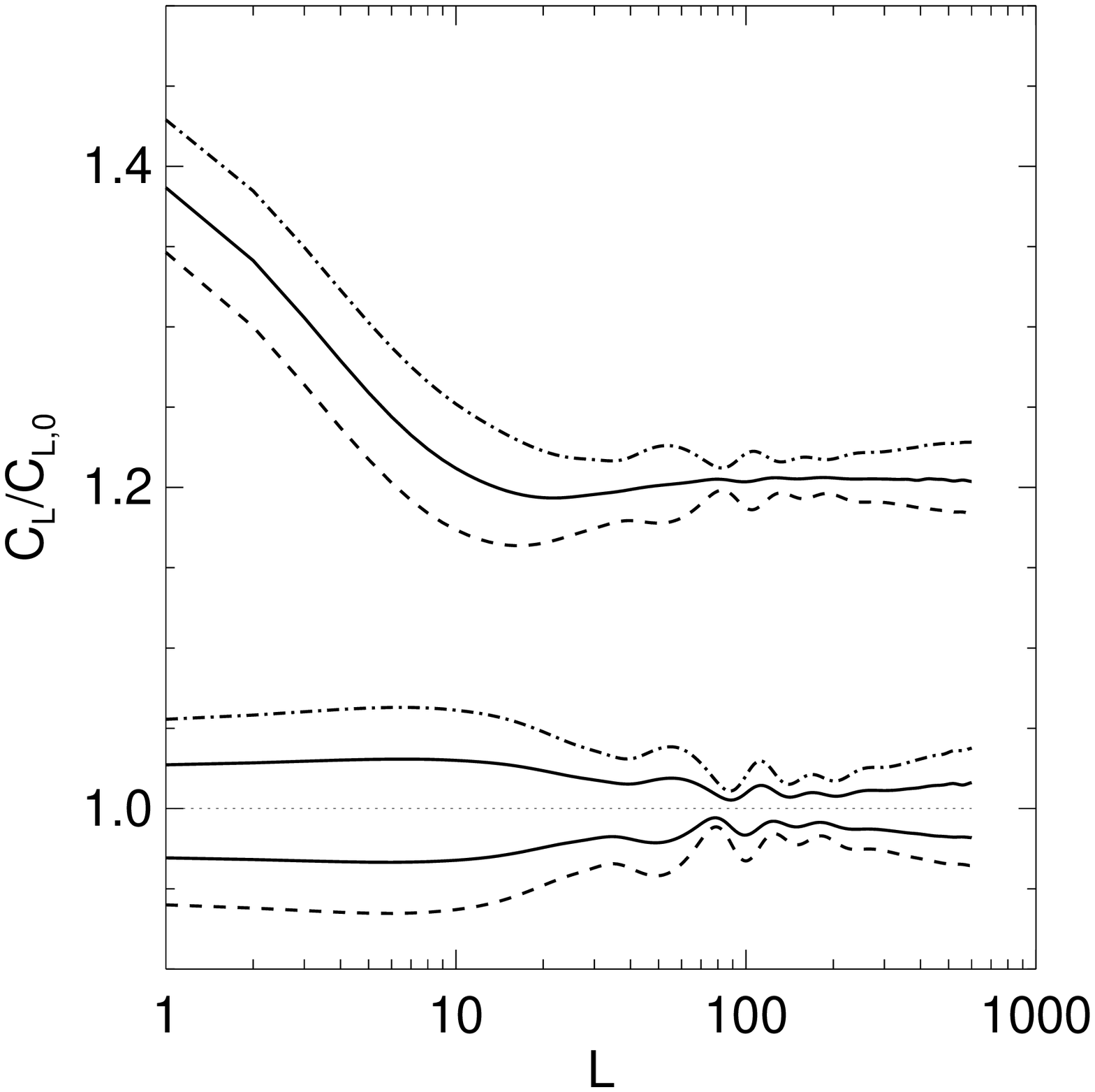}
\end{center}
\caption{The change in the angular power spectrum when the 
$z02$ redshift distribution is shifted by 
$\Delta z=0.01$ [dashed], the $z03$ redshift distribution is
shifted by $\Delta z=-0.01$ [dot-dashed], and $z02$ and 
$z03$ are shifted by $0.005$ and $-0.005$ respectively [solid].
The ratios of the angular auto power spectra are approximately 
1 ($\pm 0.05$), while the ratios of the cross correlations are approximately
1.2.
}
\label{fig:lrg_zshift}
\end{figure}

A second possibility is errors in the 
redshift distributions. To quantify this, we model possible 
redshift errors by a shift in the median redshift. An example of this
is shown in Fig.~\ref{fig:lrg_zshift} for the $z02$ and $z03$ 
slices. The figure demonstrates that shifting the median by 10\%
of the slice width can account for the discrepancies in the
cross-correlation power spectrum. Furthermore, note that the 
corrections to the auto-power spectra are $\sim 5\%$, and is principally
a multiplicative factor that is degenerate with the bias.
Finally, the above discussion also demonstrates that the 
cross-correlation spectra are able to constrain
errors in the median redshift at the percent level.

\subsection{Calibration Errors}

The final systematic effect we consider is photometric calibration errors. 
Fluctuations in the photometric calibration will select slightly different populations 
of galaxies over the entire survey region, imprinting the pattern of photometric
zeropoint errors on the derived density fluctuations. One expects
calibration errors to result in striping perpendicular to the drift scan direction (approximately
RA). These would have a characteristic scale of $\sim 0.22^{\circ}$ (the width of a camera column),
corresponding to a multipole $l \sim 800$, corresponding to smaller scales than those considered in 
this paper. The situation is further improved by the fact that the SDSS drift-scan ``strips'' are 
often broken up into several pieces with different photometric zeropoints, further reducing the
coherence length. Thus, on the angular scales used in this paper, one expects calibration errors to have 
an approximately white noise power spectrum.

\begin{figure}
\begin{center}
\leavevmode
\includegraphics[width=3.0in]{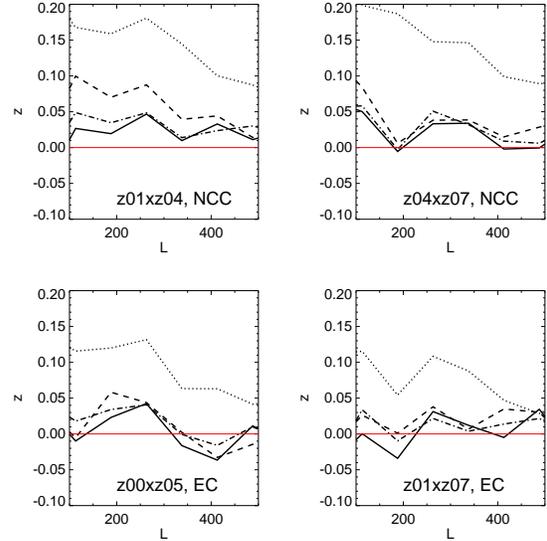}
\end{center}
\caption{The effect of calibration errors on the cross-correlation power spectrum
of non-adjacent redshift slices for a single simulation. 
The solid line shows the observed cross correlation (as the Fisher $z$-transform), 
while the dotted, dashed and dot-dashed lines show the effects of 5\%, 2\%, and 1\%
calibration errors.
}
\label{fig:simxcorr}
\end{figure}

A useful diagnostic of photometric calibration errors is the cross-correlation
between redshift slices with negligible physical overlap; calibration errors
will be common between both slices. Estimating the induced cross-correlation
requires simulations to propagate the calibration errors through the selection criteria
and photometric redshift estimation. We simulate this by perturbing the magnitude 
zeropoint of each camera column and filter separately; the resulting catalogs are then 
input into the LRG selection and photometric redshift pipelines. 

Fig.~\ref{fig:simxcorr} shows example cross-correlations for one of these simulations.
The lack of an observed cross-correlation argues for photometric calibration errors $< 2\%$,
consistent with other astrophysical tests of the calibration (D.~P.~Finkbeiner, private
communication). The effect of such errors on the autocorrelation measurements is subdominant to the
statistical errors. Note that the survey scanning 
strategy makes the large scale power spectrum relatively insensitive to $\sim 1\%$ calibration errors,
the expected calibration accuracy of the SDSS.

\section{The 3D Power Spectrum}
\label{sec:3d}

Although the above power spectra are a perfectly good representation 
of the cosmological information contained in these data, there are 
advantages to compressing these eight 2D power spectra into a single 3D power spectrum.
The first is aesthetic; given a cosmological model,
the 3D power spectrum can be directly compared to theory, 
in contrast to the 2D power spectra which involve convolutions by kernels
determined by the redshift distributions of the galaxies (that contain no 
cosmological information by themselves). 
Furthermore, the 3D power spectrum directly shows the scales probed,
and allows one to test (in a model independant manner) for features like baryonic oscillations.
Finally, the 2D power spectrum requires computing the convolution kernels, making it 
expensive to use in cosmological parameter estimations. We however emphasize that 
this is (as shown below) simply a linear repackaging of the data.

\subsection{Theory}

Inverting a 2D power spectrum to recover the 3D power spectrum
has been discussed by \cite{1998ApJ...506...64S} and \cite{2001ApJ...546....2E}.
An important detail where the two methods differ is in how they regularize the 
inversion. Since the 2D spectrum is the result of a convolution of the
3D power spectrum, it is generally not possible to reconstruct the 
3D power spectrum exactly given the 2D spectrum, and one must regularize 
the inversion. In practice, this limitation is not severe, since one would normally
estimate the power spectrum in a finite number of bands; these regularize the 
inversion if the band width approximately corresponds to the width of the 
convolution kernel. This is the solution that \cite{1998ApJ...506...64S} presents. \cite{2001ApJ...546....2E}
consider bands that have sub-kernel width, and regularize the inversion by conditioning
singular modes in an SVD decomposition. These modes are, however, given a large error,
and so contain no information.
We adopt the regularization scheme of  \cite{1998ApJ...506...64S}.
 
We start by writing the 3D power spectrum, $\Delta^{2}(k)$ as,
\be
\Delta^{2}(k) = \delta(k) \Delta^{2}_{0}(k) \,\,,
\ee
where $\delta(k)$ is the sum of step functions whose amplitudes are to be
determined, while $\Delta^{2}_{0}(k)$ is a fiducial power spectrum that describes the
shape of the power spectrum within a bin. If we now describe both
the 2D power spectrum, $C_{l}$, and and the 3D power spectrum $\delta(k)$,
as vectors of bandpowers, Eq.~\ref{eq:cl}
can be rewritten as a matrix equation,
\be
\vec{C}_{l} = {\bf W} \vec{\delta} \,\,,
\label{eq:pk_invert}
\ee
where ${\bf W}$ is the discretized convolution kernel. The solution, by 
singular value decomposition or normal equations \cite[see][15.4]{1992nrfa.book.....P}, is
\bea
{\bf C}_{\delta}^{-1} = {\bf W}^{t} {\bf C}_{cl}^{-1} {\bf W} \,\, \nonumber \\
\vec{\delta} = {\bf C}_{\delta} {\bf W}^{t} {\bf C}_{cl}^{-1} \vec{C}_{l} \,\,,
\label{eq:3dinvert}
\eea
where ${\bf C}_{cl}$ and ${\bf C}_{\delta}$ are the covariance matrices of 
$C_{l}$ and $\delta(k)$ respectively.

The above discussion glossed over a number of subtleties.
The first is extending this formalism for $N$ 2D power spectra.
If we assume that these 2D power spectra are derived from the same
3D power spectrum, one just expands $\vec{C}_{l}$ and ${\bf C}_{cl}$ to
contain all the power spectra. However, in general, the 3D power spectra
that corresponds to each of the 2D power spectra could differ both in their
bias and nonlinear evolution. For the latter, we divide $\delta(k)$ 
into two sets of bands, linear bands with $k < k_{nl}$, and nonlinear bands 
with $k \ge k_{nl}$. We then assume that the linear bands are common to all
$N$ 2D power spectra, but that there are $N$ copies of the nonlinear bands
that correspond to each of the $N$ power spectra. In what follows, we assume
that $k_{nl} = 0.1 h {\rm Mpc}^{-1}$.

Accounting for differences in bias over the different redshift slices (as
seen in Fig.~\ref{fig:lrg_bias}) is more involved. Naively adding $N$ bias
parameters to Eq.~\ref{eq:pk_invert} destroys the linearity of the system.
One might simply use the best fit values in Fig.~\ref{fig:lrg_bias}, but 
the fiducial model used might not correspond to the best fit model. We therefore
use an iterative scheme and minimize the $L^{2}$ norm of
\be
\frac{\vec{C}_{l}}{\vec{b}} - {\bf W} \vec{\delta} \,\,,
\ee
where $\vec{b}$ is a vector of the biases (squared); these biases are then 
held constant and the inversion is performed as above.

The next subtlety involves the choice of $\beta$ in order to compute the
redshift space distortions. As Fig.~\ref{fig:beta} shows, these data only weakly
constrain $\beta$, and therefore we choose to use the linear theory 
prediction for $\beta$ (more precisely for $f = b\beta$), since the redshift 
space distortions only affect the largest (and therefore most linear) scales. 

Finally, correctly combining the different redshift slices requires knowing 
the covariance between the slices. However, the power spectrum estimation in 
Sec.~\ref{sec:angpower} treats each slice independently and does not return 
the covariance between the different slices. In order to estimate the 
magnitude of this effect, we start by observing that the covariance between 
redshift slices $1$ and $2$ for multipole $l$, $C(l_{1},l_{2})$ is 
\be
C(l_{1},l_{2}) \sim 2[C^{12}_{l}]^{2}
\ee
where $C^{12}_{l}$ is the angular cross power spectrum. Using the fact that the
above relation is exact for full sky surveys, we substitute this into 
Eq.~\ref{eq:3dinvert}, and use the results with and without these redshift correlations
to scale the errors we obtain from the inversion. We discuss the validity of these
approximations below.

\subsection{Results}

\begin{figure}
\begin{center}
\leavevmode
\includegraphics[width=3.0in]{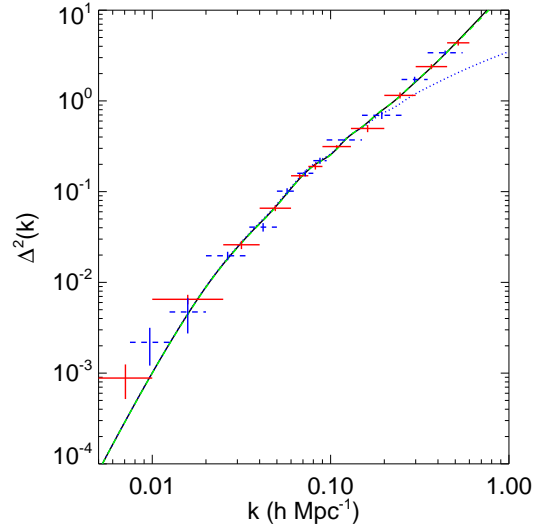}
\end{center}
\caption{The 3D power spectrum obtained by inverting the 8 2D power spectra,
normalized to the $z=0.2$ power spectrum on linear scales ($k < k_{nl}$),
and uses the $z=0.2$ bands for the nonlinear bands.
The solid and dashed lines represent binnings B1 and B2 respectively, and the 
two power spectra are consistent. Note 
that these binnings are not independent, and must not be combined for fitting.
Also shown are the nonlinear power spectrum using the \texttt{halofit} nonlinear
prescription [solid, black], the linear power spectrum 
[dotted, blue], and our suggested nonlinear prescription (see below) assuming
$Q=10.5$ [dashed, green] for our fiducial cosmology. 
}
\label{fig:3dpk_full}
\end{figure}

The result of stacking the eight 2D power spectra to
obtain a single 3D power spectrum is shown in Fig.~\ref{fig:3dpk_full}. 
Note that the inversion process yields eight 3D power spectra that differ
on scales $k > k_{nl} = 0.1 h {\rm Mpc}^{-1}$; Fig.~\ref{fig:3dpk_full} shows
the power spectrum for $z=0.2-0.25$ slice which covers the largest dynamical
range in wavenumber. Also note that the normalization of the power spectrum
is arbitrary; we normalize it to the amplitude of the power spectrum at 
$z \sim 0.2$ in the figure.
Fig.~\ref{fig:3dpk_full} shows two different binnings 
(hereafter B1 and B2)
of the power spectrum interleaved with one another; the consistency of the 
estimated power spectra demonstrates an insensitivity to the choice of 
binning.

\begin{table}
\begin{tabular}{ccccc}
\hline
$k_{min}$ & $k_{max}$ & $\Delta^{2}_{0}$ & $\delta$ & $\sigma_{\delta}$  \\
\hline
 0.005 & 0.010 &  2.8639E-04 &  2.2986E+00 &  8.7243E-01 \\
 0.010 & 0.025 &  4.4282E-03 &  1.0989E+00 &  1.1675E-01 \\
 0.025 & 0.040 &  2.1702E-02 &  8.9660E-01 &  8.2658E-02 \\
 0.040 & 0.060 &  5.3956E-02 &  9.1448E-01 &  5.8324E-02 \\
 0.060 & 0.075 &  1.0630E-01 &  1.0612E+00 &  6.0193E-02 \\
 0.075 & 0.090 &  1.5237E-01 &  9.3736E-01 &  6.0019E-02 \\
 0.090 & 0.130 &  2.3303E-01 &  1.0118E+00 &  3.2957E-02 \\
 0.130 & 0.200 &  4.4947E-01 &  1.0281E+00 &  5.4245E-02 \\
 0.200 & 0.300 &  8.5115E-01 &  1.2406E+00 &  5.0454E-02 \\

\hline
\end{tabular}
\caption{\label{tab:lrgpk1} The 3D real space power spectrum (for 
bins B1). The bands are step functions defined by $k_{min} < k < k_{max}$,
the fiducial power spectrum by $\Delta^{2}_{0}$, and the 
estimated power spectrum and errors by $\delta$ and $\sigma_{\delta}$. Note
that the full covariance matrix must be used for any detailed fitting to
these data, since different data points are anti-correlated.
}
\end{table}

\begin{table}
\begin{tabular}{ccccc}
\hline
$k_{min}$ & $k_{max}$ & $\Delta^{2}_{0}$ & $\delta$ & $\sigma_{\delta}$  \\
\hline
 0.007 & 0.013 &  7.6073E-04 &  2.0776E+00 &  7.1312E-01 \\
 0.013 & 0.020 &  3.6199E-03 &  9.4449E-01 &  2.8597E-01 \\
 0.020 & 0.035 &  1.4566E-02 &  9.7928E-01 &  8.9388E-02 \\
 0.035 & 0.050 &  3.7910E-02 &  7.7955E-01 &  7.3753E-02 \\
 0.050 & 0.065 &  7.4435E-02 &  9.9163E-01 &  6.6288E-02 \\
 0.065 & 0.080 &  1.2342E-01 &  9.4425E-01 &  5.6484E-02 \\
 0.080 & 0.095 &  1.6452E-01 &  9.7427E-01 &  6.3003E-02 \\
 0.095 & 0.150 &  2.7896E-01 &  9.6809E-01 &  2.5155E-02 \\
 0.150 & 0.250 &  5.9607E-01 &  1.0969E+00 &  4.4514E-02 \\
 0.250 & 0.350 &  1.1610E+00 &  1.1772E+00 &  5.1480E-02 \\

\hline
\end{tabular}
\caption{\label{tab:lrgpk2} Same as Table~\ref{tab:lrgpk1} except for 
bins B2.}
\end{table}

\begin{figure}
\begin{center}
\leavevmode
\includegraphics[width=3.0in]{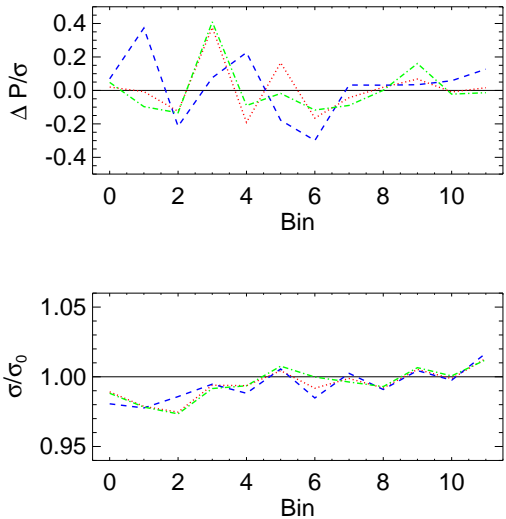}
\end{center}
\caption{(Top) The change in the the recovered power spectrum 
relative to the error, after marginalizing over a constant 
multiplicative bias, for different cosmologies/ prior power spectrum shapes.
(Bottom) The ratio of the errors relative to the fiducial case
for the same set of cosmologies/ prior power spectra.
}
\label{fig:3dpk_ratio}
\end{figure}

A second assumption necessary for the inversion is the choice of a 
cosmology to convert redshifts to distances. In principle, the consistency
between the different slices is a sensitive test of the cosmological model;
however, the errors in these data are much larger than this
effect. In order to test this, we redo the inversion with 3 different 
cosmological models, and compare the results in Fig.~\ref{fig:3dpk_ratio}
after marginalizing over the bias. Note that the changes in the power spectrum
are significantly smaller than the associated errors, while the errors in the
power spectrum remain virtually unchanged. Fig.~\ref{fig:3dpk_ratio} also
demonstrates that the inversion process does not depend on the 
particular shape of the prior power spectrum.

\begin{figure}
\begin{center}
\leavevmode
\includegraphics[width=3.0in]{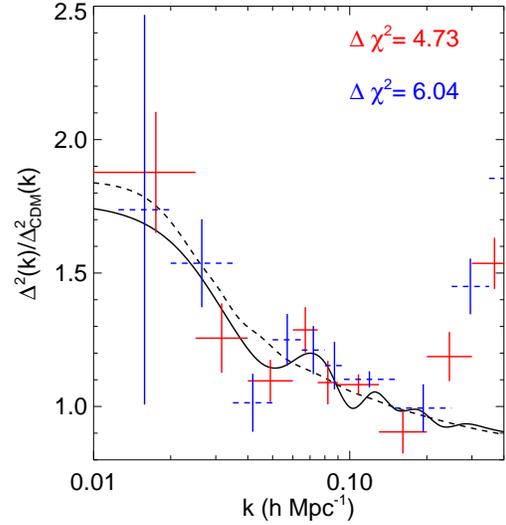}
\end{center}
\caption{The ratio of the measured power spectrum to the linear CDM power spectrum
for our fiducial cosmology (without baryons). As above, the solid and dashed lines 
represent binnings B1 and B2 respectively. Also shown is the same ratio for the
nonlinear prescription, and the ``no-wiggle'' fit to the power spectrum.
The difference in $\chi^{2}$ between these two models is shown for
the two binnings. Also note the baryonic suppression of power on large scales, and the 
rise in power due to nonlinear evolution on small scales}
\label{fig:3dpk_wiggle} 
\end{figure}

Three important features of this power spectrum are:
\begin{itemize}

\item {\it Real space power spectrum} : Since the individual angular
power spectra make no use of radial information, the 3D power 
spectrum we obtain is a real space power spectrum on small scales, avoiding 
the complications of nonlinear redshift space distortions. Note that on 
length scales much larger than the redshift slice thickness, redshift
space distortions cannot be neglected; however, the linear approximation 
discussed in Sec.~\ref{sec:theory2} will be valid on these scales. 

\item {\it Large Scale Power:} Fig.~\ref{fig:3dpk_full} shows evidence for 
power on very large ($k < 0.02 h {\rm Mpc}^{-1}$) scales. Marginalizing over
bands on smaller scales, the significance of the detection on scales 
$k < 0.01 h {\rm Mpc}^{-1}$ is $\sim 2 \sigma$, increasing to $5.5 \sigma$ for 
$k < 0.02 h {\rm Mpc}^{-1}$. Note that these scales start to probe the power spectrum 
at the turnover scale set by matter-radiation equality.

\item {\it Baryonic Oscillations:} Fig.~\ref{fig:3dpk_wiggle} shows the 3D power spectrum 
divided by a fiducial linear CDM power spectrum with zero baryonic content. The baryonic 
suppression of power on large scales, and the rise of power due to nonlinear evolution 
is clearly seen. We also see evidence for baryonic oscillations on small scales for 
both binnings, although we note that the power spectrum estimates are anti-correlated, making
a visual goodness-of-fit difficult to estimate. 

To estimate the significance of these oscillations, we compare the best fit model
obtained in the next section, with a version of the same power spectrum that has the 
baryonic oscillations edited out \citep{1998ApJ...496..605E}. The 
difference in $\chi^{2}$ for these two
models suggests a detection confidence of $\sim 2.5 \sigma$ or $\sim 95\%$, 
assuming approximately Gaussian errors. A similar result is obtained in the next 
section from cosmological parameter fits to the baryon density.

\end{itemize}

\section{Cosmological Parameters}
\label{sec:cosmo}

We defer a complete multi-parameter estimation of cosmological parameters
to a later paper, but discuss basic constraints below. We consider a
$\Lambda$CDM cosmological model, varying the matter density $\Omega_{M}$
and the baryonic fraction $\Omega_{b}/\Omega_{M}$ and fixing all other
parameters to our fiducial choices.

The principal complication to using the galaxy power spectrum for 
cosmological parameter estimation is understanding the mapping 
from the linear matter power spectrum to the nonlinear galaxy power spectrum,
both due to the nonlinear evolution of structure and 
scale-dependent bias.
We use the fitting formula proposed by \cite{2005MNRAS.362..505C}, 
\be
\frac{\Delta^{2}(k)}{\Delta^{2}_{lin}(k)} = b^{2} \frac{1+Qk^{2}}{1+A k}\,\,,
\ee
where $A=1.4$ is appropriate for a real-space power spectrum, and $b$ and 
$Q$ are two ``bias'' parameters that we add 
to the cosmological parameters we estimate. Comparing this parametrization to 
a red galaxy sample from the Millenium simulations \citep{2005Natur.435..629S},
shows that this parametrization correctly describes the effects of scale
dependent bias and nonlinear evolution up to wavenumbers 
$k \sim 0.5 h {\rm Mpc}^{-1}$ (Volker Springel, private communication).
We fit the data to $k=0.3  h {\rm Mpc}^{-1}$.

\begin{figure}
\begin{center}
\leavevmode
\includegraphics[width=3.0in]{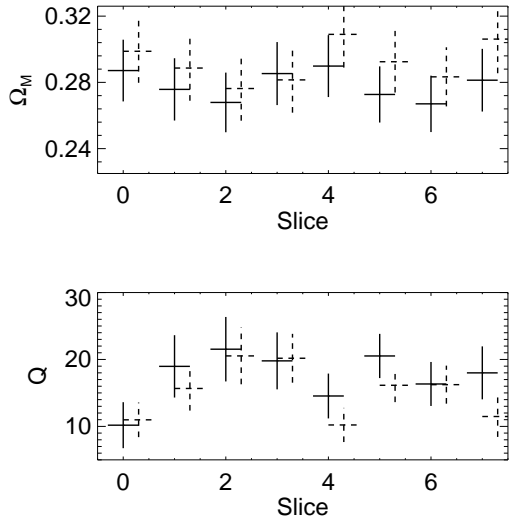}
\end{center}
\caption{Measurements of $\Omega_{M}$ (upper) and $Q$ (lower) 
for each of the 
eight 3D power spectra. The solid
lines use binning B1, while the displaced dashed lines 
use binning B2. Note that $\Omega_{M}$ is 
insensitive to the redshift slice used, while $Q$ depends 
sensitively on the particular choice of slice.}
\label{fig:cosmo2d_omq_zslice}
\end{figure}

A second complication is that the inversion procedure of the previous
section only combines wavenumbers $< 0.1 h {\rm Mpc}^{-1}$; fitting the
data beyond this requires choosing one of the eight redshift slices. 
In order to decide which slice to use, we estimate $\chi^{2}$ on grids
varying $\Omega_{M}$, $Q$, and $b$ for each of the eight 3D power
spectra. We fix the baryonic density to $\Omega_{b}=0.05$, although 
allowing it to vary does not change the results. 

The best fit values for $\Omega_{M}$ and $Q$ (marginalizing over
the other parameters), for each of the eight redshift slices are
shown in Fig.~\ref{fig:cosmo2d_omq_zslice}. We note that 
$\Omega_{M}$ and its error is insensitive to the choice of
redshift slice, although $Q$ depends on the particular redshift
slice used. This is due to the fact that $\Omega_{M}$ is constrained
by the location of the turnover in the power spectrum, and the 
shape of the power spectrum in the linear regime,  while $Q$ depends
on the power spectrum beyond $0.1 h {\rm Mpc}^{-1}$. In what follows,
we use the redshift slice corresponding to photometric redshifts 
between 0.45 and 0.50, as it corresponds to the
median redshift of the full sample. However, we emphasize that all
results below, except for the ``nuisance'' bias parameters, are insensitive 
to this particular choice.

\begin{figure*}
\begin{center}
\leavevmode
\includegraphics[width=6.0in]{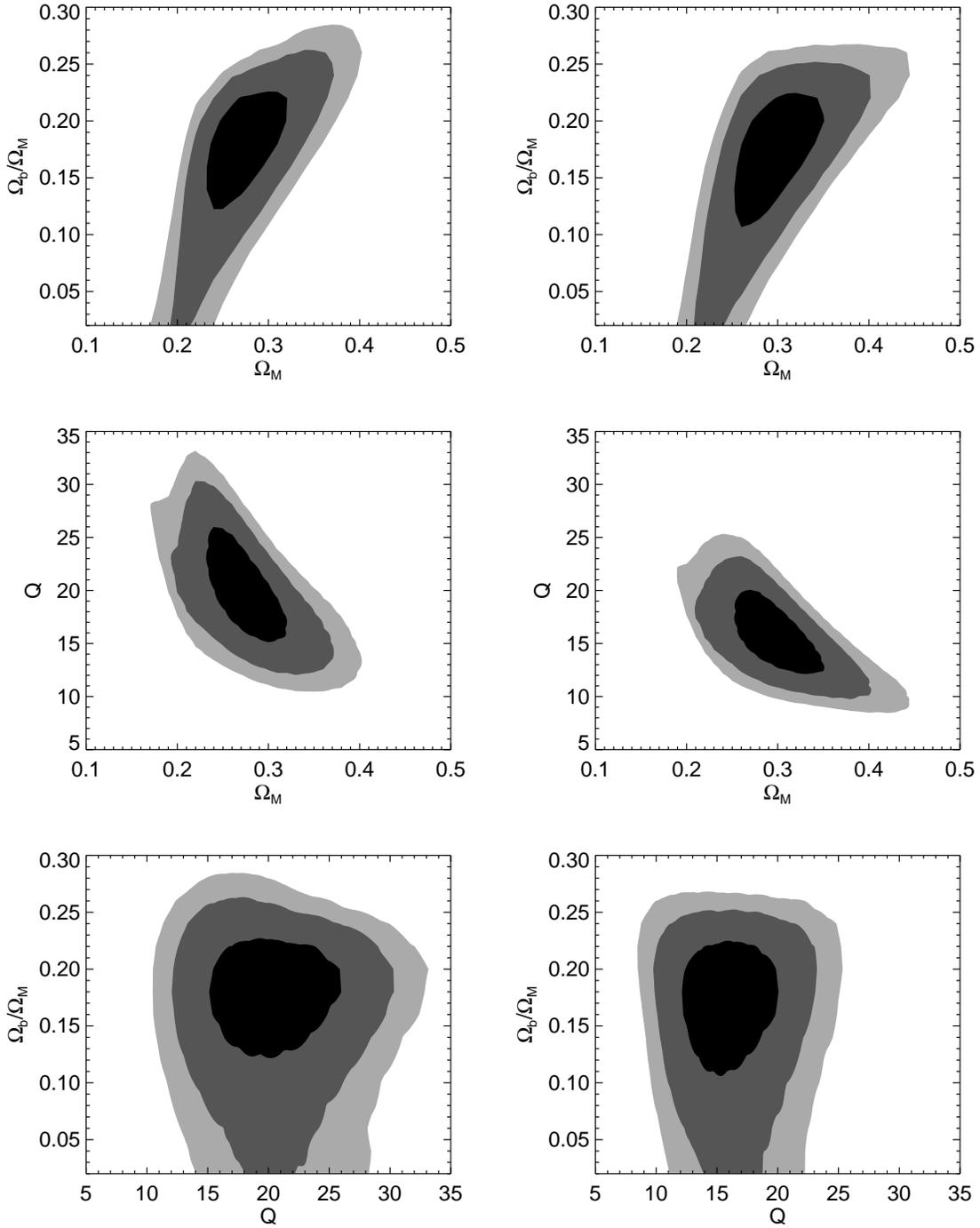}
\end{center}
\caption{Joint 2D likelihood distributions for $\Omega_{M}$, $Q$ and 
$\Omega_{b}/\Omega_{m}$, fixing $h=0.7$ and $n=1$, and marginalizing over 
the galaxy bias. The contours show $\Delta \chi^{2}=2.3,6.17$ and $9.21$.
The left column panels use binning B1, while the
right column panels use B2.  We truncate fitting at $k = 0.3 h {\rm Mpc}^{-1}$ (using the
midpoint of the bins). Note that the two binnings are consistent with each
other, with the B2 binning providing slightly tighter constraints.
}
\label{fig:cosmo2d}
\end{figure*}

Fig.~\ref{fig:cosmo2d} shows 2D projections of the $(\Omega_{M}, \Omega_{b}/\Omega_{M},
Q)$ parameter likelihood space; the multiplicative bias $b$ is marginalized over. 
The minimum $\chi^{2}$ values are $5.99$ and $6.94$ (bins B1 and B2, respectively), for 
$5$ and $6$ degrees of freedom, consistent with a reduced $\chi^{2}$ of 1 per degree
of freedom.
Bins B1 and B2 give consistent values for the cosmological parameters; 
B2 constrains all parameters (especially $Q$) better than B1 because of the 
extra binning and the larger $k$ range probed. We note that $Q$ is correlated with 
$\Omega_{M}$, since both $\Omega_{M}h$ and $Q$ determine the broad shape of the 
power spectrum. An important consequence of this degeneracy is that an accurate
estimation of $\Omega_{M}$ and its error requires varying $Q$; fixing or restricting
$Q$ will result in a biased $\Omega_{M}$ and an underestimation of its errors. 

\begin{figure}
\begin{center}
\leavevmode
\includegraphics[width=3.0in]{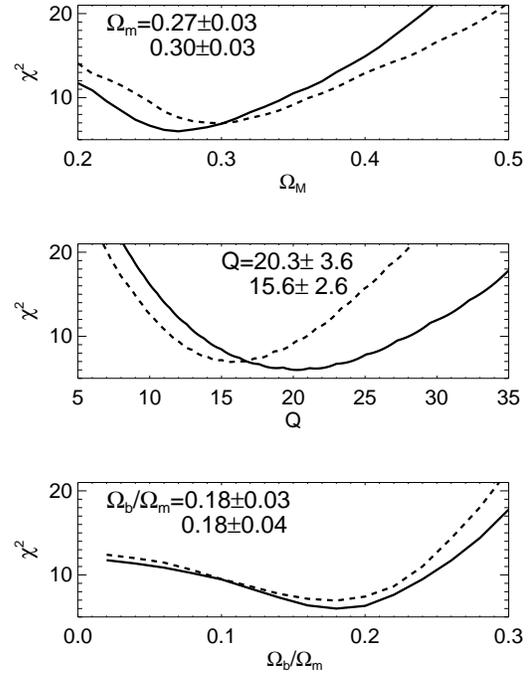}
\end{center}
\caption{The likelihood distributions for $\Omega_{M}$, $Q$ and $\Omega_{b}/\Omega_{M}$.
The solid line uses binning B1, while the dashed line uses B2. As 
in Fig.~\ref{fig:cosmo2d}, we truncate at $k = 0.3 h {\rm Mpc}^{-1}$. Also shown are
the best fit values and $1 \sigma$ errors for bins B1 (top) and B2 (bottom).
}
\label{fig:cosmo_om}
\end{figure}

Fig.~\ref{fig:cosmo_om} shows the 1D likelihoods for  $(\Omega_{M}, \Omega_{b}/\Omega_{M},
Q)$, marginalizing over all other parameters; the binnings are again seen to be 
consistent. The likelihood for $\Omega_{b}/\Omega_{M}$ also allows to estimate the 
significance of the detection of baryonic features in the power spectrum. The 
difference in $\chi^{2}$ between the best fit model and the zero-baryon case is 
$5.75$ and $6.4$ for bins B1 and B2 respectively, suggesting a $2.5\sigma$ detection 
consistent with the model independent estimates made in the previous section.
The significance of this result is similar to the results from the 2dFGRS \citep{2005MNRAS.362..505C}, but is 
weaker than the detection in the spectroscopic LRG sample \citep{2005ApJ...633..560E}.

Summarizing these results, we have
\begin{itemize}
\item For bins B1 : 
\bea
\Omega_{M} = 0.27 \pm 0.03 \nonumber \\
\Omega_{b}/\Omega_{M} = 0.18 \pm 0.03 \nonumber \\
Q = 20.3 \pm 3.6 
\eea
\item For bins B2 :
\bea 
\Omega_{M} = 0.30 \pm 0.03 \nonumber \\
\Omega_{b}/\Omega_{M} = 0.18 \pm 0.04 \nonumber \\
Q = 15.6 \pm 2.6 
\eea
\end{itemize}

In light of the recent WMAP results \citep{2006astro.ph..3449S}, it is interesting 
to understand how the above reults change if we deviate from a scale-invariant 
primordial spectrum. Minimizing $\chi^{2}$ over $\Omega_{M}$, $\Omega_{b}/\Omega_{M}$, 
and $Q$ assuming $n=0.95$, we find that (for bins B2), 
\bea
\Omega_{M} = 0.31 \pm 0.03 \nonumber \\
\Omega_{b}/\Omega_{M} = 0.16 \pm 0.04 \nonumber \\
Q = 16.3 \pm 2.8 \,\,. 
\eea
Reducing $n$ (while keeping $\sigma_{8}$ fixed) boosts the power on 
large scales, but suppresses it on small scales. This results in a better fit
on large scales, and a worse fit on small scales. To compensate for this, 
the best fit value of $\Omega_{b}/\Omega_{M}$ decreases (reducing Silk damping) 
while $Q$ increases, 
boosting the power back up on small scales, while leaving the large scale
power spectrum unchanged. The minimum $\chi^{2}$ is marginally worse ($7.24$)
than the scale invariant case. Note however that all the parameters are
within the 1-$\sigma$ errors of those obtained assuming scale invariance.

\subsection{Distance to $z=0.5$}

\begin{figure}
\begin{center}
\leavevmode
\includegraphics[width=3.0in]{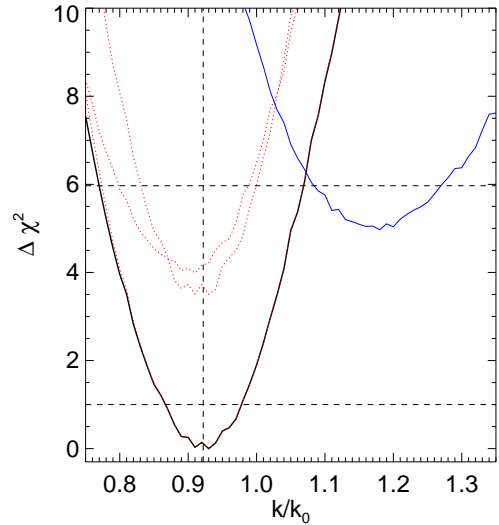}
\end{center}
\caption{The solid line shows the likelihood of the ratio of the fiducial distance to $z=0.5$
to the observed distance $k/k_{0}$, marginalizing over $Q$ and the galaxy bias, but fixing 
$\Omega_{b}h^{2}=0.0223$, $\Omega_{M}h^{2}=0.127$ and $Q=15.6$. The dotted lines show slices
through the 2D likelihood distribution of $Q$ and $k/k_{0}$ at $Q=16.5$, $19.5$ and $22.5$
(from left to right). Note that the $Q$ and $k/k_{0}$ are approximately 
orthogonal directions; varying the nonlinear correction doesn't change the distance scale.
The best fit value of $0.92 \pm 0.06$ is shown by the dashed lines. 
The thin solid line shows the $k/k_{0}$ likelihood for a negligible baryonic fraction; the
distance constraint degrades to a $10\%$ measurement.
}
\label{fig:lrg_distance}
\end{figure}

A potential application of the galaxy power spectrum is as a standard ruler. The two 
features of interest, the turnover and the baryon oscillations are determined by the physical
baryon and matter densities - $\Omega_{b}h^{2}$ and $\Omega_{M}h^{2}$. Both of these are
precisely determined by the peak structure of the CMB power spectrum. Therefore, in order
to understand the sensitivity of the current measurements as standard rulers, we fix 
$\Omega_{b}h^{2}=0.0223$ and $\Omega_{M}h^{2}=0.127$, and vary $Q$ and the comoving distance. 
In general, one would need to vary the comoving distance to each of the 8 redshift slices
and recompute the power spectrum. However, given the S/N of the baryonic oscillations and
turnover in these data, we simply translate the 3D power spectrum in $k$ with reference to our
fiducial cosmology at the median redshift of the slice $k_{0}$. The likelihood is in 
Fig.~\ref{fig:lrg_distance}; these data can constrain the distance to $z=0.5$ to $6\%$.
Note that this is for a fixed value of $\Omega_{M}h^{2}$. Assuming a $10\%$ uncertainty in 
$\Omega_{M}h^{2}$ from current CMB measurements results in a $\sim 2.5\%$ uncertainty 
in the sound horizon, 
increasing the distance error to $6.5\%$. This must be compared to the $5\%$ measurement
of the distance to $z\sim 0.35$ measured by the spectroscopic LRG sample. 

Equally interesting is that $Q$ and $k/k_{0}$ are orthogonal; the distance measurement does 
not change for different values of the nonlinear correction. This highlights an important 
property of baryon oscillations as a distance measurement - it is relatively insensitive to 
the nonlinearity corrections that affect the galaxy power spectrum.

We would also like to understand the fraction of the distance constraint from baryonic
oscillations as opposed to the power spectrum shape. Fig.~\ref{fig:lrg_distance} also 
shows the likelihood for a model with a negligible baryonic fraction; the distance accuracy
degrades to $10\%$, suggesting that most of the constraint comes from the oscillations.

\section{Discussion}
\label{sec:discuss}

\subsection{Principal Results}

We have measured the 3D clustering power spectrum of luminous 
red galaxies using the SDSS photometric survey. 
The principal results of this analysis are summarized below.
\begin{itemize}

\item {\it Photometric redshifts:} This analysis demonstrates the feasibility
of using multi-band imaging surveys with well calibrated photometric 
redshifts as a probe of the large scale structure of the Universe.
Accurate photometric redshifts are
critical to being able to narrow the range of physical scales that correspond
to the clustering on a particular angular scale, and thereby estimate the 
3D power spectrum.

\item {\it Largest cosmological volume:} Using photometric redshifts allowed 
us to construct a uniform sample of galaxies between 
redshifts $z=0.2$ to $0.6$. This probes a cosmological volume of $\sim 1.5 
h^{-3} {\rm Gpc}^{3}$, making this the largest cosmological volume 
ever used for a galaxy clustering measurement.  The large volume allows us
to measure power on very large scales, yielding a $\sim 2 \sigma$ detection 
of power for $k < 0.01 h {\rm Mpc}$, increasing in significance to 
$\sim 5.5 \sigma$ for $k < 0.02 h {\rm Mpc}$.

\item {\it Real Space Power Spectrum:} This power spectrum is intrinsically a 
real space power spectrum, and is unaffected by redshift space distortions
on scales $k > 0.01 h {\rm Mpc}^{-1}$. This obviates any need to model redshift 
space distortions in the quasi-linear regime, allowing for a more direct
comparison to theoretical predictions.

\item {\it Baryonic Oscillations:} The 3D power spectrum shows evidence for 
baryonic oscillations at the $\sim 2.5 \sigma$ confidence level, both in the 
shape of the 3D power spectrum, as well as fits of the baryonic density.
We emphasize that this is only possible in the stacked 3D power spectrum,
and therefore relies on accurate photometric redshift distributions.

\item {\it Cosmological Parameters:} The large volume and small statistical
errors of these data constrain both the normalization and scale 
dependence of the galaxy bias. Using a functional form for the scale dependence
of the bias motivated by N-body simulations, we fit for the matter density
and baryonic fraction jointly, and obtain $\Omega_{M} = 0.30 \pm 0.03$ and 
$\Omega_{b}/\Omega_{M} = 0.18 \pm 0.03$.  

\end{itemize}

\subsection{Using these results}

For cosmological parameter analyses, we 
recommend directly using the 3D power spectra (binning B2), fitting both the 
galaxy bias ($b$) and its scale-dependence ($Q$) to $k=0.3 h {\rm Mpc}^{-1}$. 
Electronic versions of all the power spectra, and covariance matrices used in this
paper will be made publically available. 
In addition, a simple \texttt{FORTRAN} subroutine that returns $\chi^{2}$ given an input power 
spectrum will also be made public.

\subsection{Comparison with other results}

\begin{figure}
\begin{center}
\leavevmode
\includegraphics[width=3.0in]{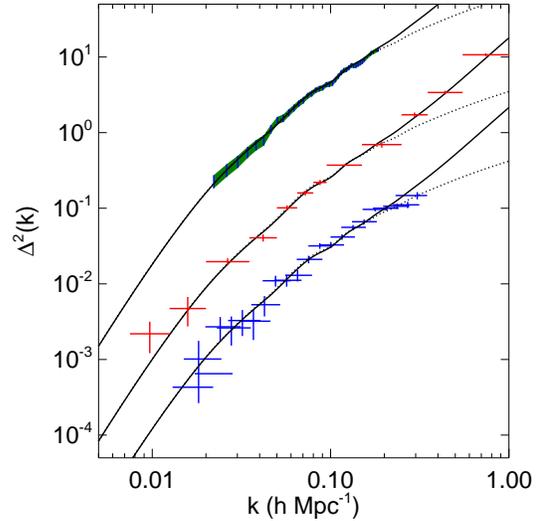}
\end{center}
\caption{Comparison between power spectra from the SDSS MAIN spectroscopic survey (bottom), 
2dFGRS $P(k)$ (top), and the photometric SDSS LRGs (binning B2) from this work (middle). The 
normalizations are arbitrary, and have been simply chosen to separate the three 
power spectra. Adjacent LRG $P(k)$ errors are anticorrelated, while the 
2dFGRS errors are strongly correlated. The dotted line shows the linear power spectrum 
for our fiducial cosmology, whereas the solid line is the \texttt{halofit} presciption
for the nonlinear power spectrum. Note that the LRG power spectrum fits the nonlinear
power spectrum to $k \sim 1 h {\rm Mpc}^{-1}$.
}
\label{fig:3dpk_compare}
\end{figure}

Fig.~\ref{fig:3dpk_compare} compares the LRG power spectrum (B2 binning), 
with the power spectrum
obtained from the SDSS MAIN spectroscopic survey \citep{2004ApJ...606..702T} and 
the 2dFGRS \citep{2005MNRAS.362..505C}; these three samples will be 
referred to as LRG, MAIN, and 2dF throughout this section. The solid and dotted lines show our 
nonlinear and linear fiducial power spectrum. Note that the normalization is arbitrary,
and that we have not attempted to deconvolve the 2dF window function.

The two principal differences between these surveys
and the data presented here is the volume probed, and the density of objects.
As both the MAIN and 2dF are at low redshifts (median $z \sim 0.1$), \
the volume probed is 
$< 0.05 h^{-3} {\rm Gpc}^{3}$, whereas our sample probes $1.5 h^{-3} {\rm Gpc}^{3}$
(at a median redshift of $z \sim 0.5$)
allowing us to measure the largest scales with smaller statistical errors, even 
with crude redshift estimates. This is clearly evident from Fig.~\ref{fig:3dpk_compare},
where the LRG power spectrum extends to smaller $k$ than either of the other 
two power spectra.

On small scales, we again emphasize that the LRG power spectrum is naturally a real space 
power spectrum, and is unaffected by redshift space distortions. By contrast, the 
2dF $P(k)$ is in redshift space, and the MAIN $P(k)$ which involves 
attempting to correct for linear redshift space distortions. 
Note that the SDSS $P(k)$ 
falls below the nonlinear power spectrum at $k \sim 0.3 h {\rm Mpc}^{-1}$, 
in line with the simulation results of \cite{2004ApJ...606..702T} that motivated the discarding of 
$k>0.2 h {\rm Mpc}^{-1}$ data from the cosmological parameter analysis. This is a
manifestation of nonlinear redshift distortions, which are
particularly important given recent results that 
suggest that redshift distortions go nonlinear on larger scales than previously 
anticipated \citep{2006MNRAS.366.1455S}.

\begin{table}
\begin{tabular}{lc}
\hline
Survey & $\Omega_{M}$ \\
\hline
SDSS MAIN & $0.297\,(+0.0219,-0.0196)$ \\
2dFGRS & $0.271\, (+0.021,-0.0187)$ \\
SDSS LRG (B1) & $0.260\,(+0.0111,-0.0102)$ \\
SDSS LRG (B2) & $0.286\,(+0.0119,-0.0111)$ \\
\hline
\end{tabular}
\caption{\label{tab:compare_omegam} The best fit values for $\Omega_{M}$
assuming $\Omega_{b} h^{2}=0.024$, $h=0.72$ and a scale invariant initial 
perturbation spectrum. We use the best fit nonlinear prescription suggested
by the respective authors to fit the power spectrum to $k = 0.2 h {\rm Mpc}^{-1}$.
The numbers in parentheses are the upper and lower $1-\sigma$ errors.  
}
\end{table}

Each of the three power spectra are consistent with the overall shape of the fiducial 
power spectrum, suggesting that they are consistent with each other. In order to make
this precise and to compare statistical power, we fit for $\Omega_{M}$ and the galaxy bias
assuming $\Omega_{b} h^{2}=0.024$, $h=0.72$ and a scale invariant initial 
perturbation spectrum. To ensure a fair comparison, we fit all power spectra to 
$k = 0.2 h {\rm Mpc}^{-1}$ using the best fit prescription for 
nonlinearity suggested by the authors. The results are summarized in Table~\ref{tab:compare_omegam}; 
we find that all three power spectra yield consistent values for $\Omega_{M}$. 
The LRG power spectrum, however, reduces the error by a factor of $\sim 1.75$ compared 
with previous results.

On the other hand, the SDSS LRG spectroscopic sample is similar to this sample. 
The effective spectroscopic LRG volume is $0.75 h^{-3} {\rm Gpc}^{3}$ at a median 
redshift of $z \sim 0.35$. However, the spectroscopic LRGs are sparser,
with shot noise responsible for approximately half the statistical error on all scales.   
One can compare the S/N of the two samples as follows - since we 
are only using auto-correlations of the redshift slices and
are ignoring correlations between different redshift slices, 
we are losing a factor $\sim 7$ in the number of modes (most
of the remaining cosmological information is contained in adjacent redshift slices).
We however gain a factor $\sim 2$ from the increased volume, and another factor 
$\sim 2$ from the higher spatial density of objects, suggesting that the 
SDSS spectroscopic LRG sample would be a factor of $\sim 2$ greater in S/N 
than the photometric sample. This is borne out by the fact that the spectroscopic 
sample detects baryonic oscillations with $\Delta \chi^{2} = 11.7$, while the 
photometric sample has $\Delta \chi^{2} = 6.04$, about a factor of $2$ smaller.
Note that this analysis breaks down both on the largest scales (where the 
photometric survey has more leverage because of the greater volume), and on scales
smaller than the redshift errors (where the spectroscopic sample resolves more modes).
In principle, one could gain further by using the cross correlations between different 
redshift slices. However, as seen in Fig.~\ref{fig:lrg_zshift}, this is very sensitive 
to errors in the tails of the redshift distribution.

We can also compare our cosmological results with those obtained from the third year
CMB temperature and polarization measurements from the WMAP satellite \citep{2006astro.ph..3449S}. 
The WMAP error on $\Omega_{M}$ is dominated by the error on the Hubble constant; they obtain
$\Omega_{M}=0.26^{+0.01}_{-0.03}$, compared with our estimate of $\Omega_{M}=0.30 \pm 0.03$.
They also obtain $\Omega_{b}/\Omega_{M} = 0.17$, compared with $\Omega_{b}/\Omega_{M} = 0.18 \pm 0.04$. 
Note that WMAP favours a primordial scalar spectral index of $n \sim 0.94$; using this instead
of scale invariance reduces our estimate of $\Omega_{b}/\Omega_{M}$ to $0.16 \pm 0.04$, 
while increasing $\Omega_{M}=0.31 \pm 0.03$. We also emphasize that the errors are not 
directly comparable, since our analysis uses stricter priors. It is, however, important and
encouraging to note that we obtain consistent results with a completely independent dataset.

\subsection{Future Directions}

We conclude with a discussion of the future prospects for photometric surveys. As of
this writing, the SDSS has imaged twice the area used in this paper, potentially 
reducing the errors by a factor of $\sqrt{2}$. In addition, there are a number of 
imaging surveys planned for the near and distant future, the Pan-STARRS 
\footnote{\texttt{http://pan-starrs.ifa.hawaii.edu}} and LSST\footnote{\texttt{www.lsst.org}}
being two notable examples. Both of these surveys will ultimately cover about three
times the final SDSS area to a much greater depth, further increasing the 
volume probed.

Baryonic oscillations are also now emerging as an important tool to constrain the
properties of dark energy. The tradeoff between photometric and spectroscopic
approaches to their measurement is simple - photometric surveys require wide field 
($> 10,000 {\rm deg}^{2}$)
multi-band imaging surveys, whereas spectroscopic surveys require large multi-object spectrographs.
Both of these approaches are being actively developed, and the prudent 
approach would be to pursue both, using the results from one to inform the other. 
It is worth emphasizing that wide-field imaging surveys are an essential prerequisite
for the other approaches (with very different systematic errors) 
to understanding dark energy, namely supernovae and weak lensing,
suggesting a synergy between these techniques.

Given the efforts underway to plan the next generation of surveys, it is timely to 
compare the precision of the distance measurement we obtain with the fitting 
formulae of \cite{2006MNRAS.365..255B}. Substituting our survey parameters into their
photometric fitting formula, assuming a median redshift of $z\sim 0.5$ and a photometric 
redshift error $\sigma_{z} \sim 100 h^{-1} {\rm Mpc}$ (corresponding to the redshift error
at $z\sim0.5$), we estimate a distance error of $7\%$ 
as compared with the actual $6\%$ error obtained. Note that \cite{2006MNRAS.365..255B}
only use the oscillation to determine the distance, whereas we use the 
entire power spectrum.

We can now estimate the potential sensitivity of the next generation of surveys. 
Assuming a straw-man survey of $20,000 {\rm deg}^{2}$ with a median redshift of $z \sim 0.8$,
and photometric redshift errors of $\sim 50 h^{-1} {\rm Mpc}$, we find a factor of $\sim 5$
improvement in the distance measurement, yielding a $\sim 1\%$ measurement, the current
benchmark for dark energy surveys.
Note that this is a conservative estimate, since the photometric redshift accuracies assumed
have already been achieved with the SDSS. 

In order to do so, there are a number of challenges that must be overcome, in addition 
to the brute force observational effort required. The first is technical -
this work relied heavily on having accurate, well calibrated photometric redshifts. 
Demonstrating that this is possible at higher redshifts, and calibrating the redshift 
errors is essential. The second challenge is theoretical - in order to optimally
use galaxy clustering for cosmology, we will now need to understand the connections 
between the physics of galaxy formation and the observed clustering of galaxies. The 
hope is that the interplay between the two would result in a more complete 
cosmological model. 

We thank Lloyd Knox, Eric Linder, Yeong Loh, Taka Matsubara, David Weinberg
and Martin White for useful discussions. We also thank the 2SLAQ collaboration for 
measuring the spectroscopic redshifts used to calibrate the photometric redshift algorithm.

Funding for the SDSS and SDSS-II has been provided by the Alfred P. Sloan Foundation, the Participating 
Institutions, the National Science Foundation, the U.S. Department of Energy, the National Aeronautics 
and Space Administration, the Japanese Monbukagakusho, the Max Planck Society, and the Higher Education 
Funding Council for England. The SDSS Web Site is http://www.sdss.org/.

The SDSS is managed by the Astrophysical Research Consortium for the Participating Institutions. The Participating 
Institutions are the American Museum of Natural History, Astrophysical Institute Potsdam, University of Basel, 
Cambridge University, Case Western Reserve University, University of Chicago, Drexel University, Fermilab, 
the Institute for Advanced Study, the Japan Participation Group, Johns Hopkins University, the Joint 
Institute for Nuclear Astrophysics, the Kavli Institute for Particle Astrophysics and Cosmology, the Korean 
Scientist Group, the Chinese Academy of Sciences (LAMOST), Los Alamos National Laboratory, the Max-Planck-Institute 
for Astronomy (MPIA), the Max-Planck-Institute for Astrophysics (MPA), New Mexico State University, Ohio 
State University, University of Pittsburgh, University of Portsmouth, Princeton University, the United States 
Naval Observatory, and the University of Washington.

The 2SLAQ Redshift Survey was made possible through the dedicated efforts of the
staff at the Anglo-Australian Observatory, both in creating the 2dF
instrument and supporting it on the telescope.

\bibliography{biblio,preprints}
\bibliographystyle{mnras}

\end{document}